\documentclass{aa}  

\usepackage{graphicx}
\usepackage{txfonts}
\usepackage[colorlinks]{hyperref}
\usepackage[dvipsnames]{xcolor}
\usepackage[printonlyused]{acronym}
\usepackage{ulem}

\acrodef{FRMS}{fast-rotating massive stars}
\acrodef{SMS}{super massive stars}
\acrodef{GC}{Globular clusters}
\acrodef{1P}{first population}
\acrodef{2P}{second population}
\acrodef{MS}{main-sequence}
\acrodef{ZAMS}{zero-age main-sequence}
\acrodef{RGB}{red giant branch}
\acrodef{RG}{red giant}
\acrodef{HB}{horizontal branch}
\acrodef{CHeB}{core helium-burning}
\acrodef{AGB}{asymptotic giant branch}
\acrodef{TP-AGB}{thermal pulsating asymptotic giant branch}
\acrodef{WD}{white dwarf}
\acrodef{HR}{Hertzsprung-Russell}
\newcommand{\Msun}{\mbox{M$_{\odot}$}}

\newcommand{\Teff}{\mbox{$T_{\rm eff}$}}
\newcommand{\feh}{\mbox{\rm [{\rm Fe}/{\rm H}]}}
\newcommand{\afe}{$\mathrm{[\alpha/Fe]}$}

\begin{document}

   \title{He-enriched \textsc{starevol} models for globular cluster multiple populations}

   \subtitle{Self-consistent isochrones from ZAMS to the TP-AGB phase}

   \author{G. Costa
          \inst{1}
          % \fnmsep\thanks{guglielmo.costa.astro@gmail.com}
          \and
          T. Dumont\inst{2}
          \and
          A. Lançon\inst{3}
          \and
          A. Palacios\inst{4}
          \and
          C. Charbonnel\inst{5,6}
          \and 
          P. Prugniel\inst{1}
          \and
          S. Ekstrom\inst{5}
          \and \\
          C. Georgy\inst{5}
          \and
          V. Branco\inst{3,7}
          \and
          P. Coelho\inst{7}
          \and
          L. Martins\inst{8}
          \and
          S. Borisov\inst{5}
          \and
          K. Voggel\inst{3}
          \and
          W. Chantereau\inst{3}
          }

   \institute{Univ Lyon, Univ Lyon1, ENS de Lyon, CNRS, Centre de Recherche Astrophysique de Lyon UMR5574, F-69230 Saint-Genis-Laval, France. 
              \email{guglielmo.costa.astro@gmail.com}
         \and
            University of Strasbourg, CNRS, IPHC UMR 7178, F-67000 Strasbourg, France
        \and 
            Observatoire Astronomique de Strasbourg, Université de Strasbourg, CNRS, UMR 7550, 11 rue de l’Université, F-67000 Strasbourg, France
        \and
            LUPM, Université de Montpellier, CNRS, Place Eugène Bataillon, 34095 Montpellier, France
        \and
            Department of Astronomy, University of Geneva, Chemin Pegasi 51, 1290 Versoix, Switzerland
            \and 
            IRAP, CNRS UMR 5277 \& Universit\'{e} de Toulouse, 14 avenue Edouard Belin, 31400 Toulouse, France 
        \and 
            Universidade de São Paulo, IAG, Rua do Matão, 1226, 05508-090, Sao Paulo, SP, Brazil.
        \and
            NAT - Universidade Cidade de São Paulo, Rua Galvão Bueno, 868, 01506-000, São Paulo, SP, Brazil.
        }

   \date{Received XXX, XXXX; accepted XXXXX XX, XXXX}

% \abstract{}{}{}{}{} 
% 5 {} token are mandatory
 \hypersetup{
    linkcolor=blue,
    citecolor=blue,
    filecolor=magenta,      
    urlcolor=blue
}
\abstract{
    A common property of globular clusters (GCs) is to host multiple populations characterized by peculiar chemical abundances.
    Recent photometric studies suggest that the He content could vary between the populations of a GC by up to $\Delta$He $\sim$ 0.13, in mass fraction. The initial He content impacts the evolution of low-mass stars by ultimately modifying their lifetimes, luminosity, temperatures, and, more generally, the morphology of post-RGB evolutionary tracks in the Hertzsprung-Russell diagram.
    We present new physically accurate isochrones with different initial He enrichments and metallicities, with a focus on the methods implemented to deal with the post-red giant branch phases. The isochrones are based on tracks computed with the stellar evolution code \textsc{starevol} for different metallicities (Z = 0.0002, 0.0009, 0.002, and 0.008) and with a different He enrichment (from 0.25 to 0.6 in mass fraction).  We describe the effect of He enrichment on the morphology of the isochrones and we tested these by comparing the predicted number counts of horizontal branch and asymptotic giant branch stars with those of selected GCs.
    Comparing the number ratios, we find that our new theoretical ones agree with the observed values within $1\sigma$ in most cases. The work presented here sets the ground for future studies on stellar populations in GCs, in which the abundances of light elements in He-enhanced models will rely on different assumptions for the causes of this enrichment. The developed methodology permits the computation of isochrones from new stellar tracks with non canonical stellar processes. The checked number counts ensure that, at least in this reference set, the contribution of the luminous late stages of stellar evolution to the integrated light of a GC is represented adequately.
    }


   \maketitle
%
%-------------------------------------------------------------------

\section{Introduction}
\label{sec:intro}

\ac{GC} are old ($>$ 9 Gyr) and dense agglomerates of stars, present in the bulge, thick disk, and halo of the Milky Way and other galaxies.
A common property shared between \ac{GC} is the presence of multiple populations, which witnesses signs of early chemical evolution in the cluster.
Multiple populations have been found in \ac{MS}, \ac{RGB}, \ac{HB} and \ac{AGB} stars of a large sample of clusters \citep[e.g.][]{DAntona2005, Piotto2007, Carretta2010, Lind2011, Campbell2013, Wang2016, Wang2017, Milone2017, Tailo2020, Lagioia2021}.
Spectroscopic and photometric studies have revealed the different chemical compositions of the various stellar populations, which are mainly associated with the abundance of light elements \citep[from He to Sc, e.g., ][]{Carretta2015, Gratton2019, Carretta2021}.

Stars in GCs that share the same light element pattern with field stars with the same [Fe/H] are the so-called \ac{1P}\footnote{It is worth noticing that 1P could be more complex than first thought. In fact, as suggested in the study by \citet{Marino2019}, 1P stars in GCs show a potential metallicity spread}.
In contrast, stars enriched in He, N, and Na and depleted in C and O with respect to field stars are classified as \ac{2P}.
We refer the reader to \citet{Gratton2019}, \citet{Cassisi2020}, and \citet{Milone2022} for recent reviews on multiple populations in GCs.
Multiple populations in GCs have also been found in the faintest part of the \ac{MS} by using data from \textit{Hubble Space Telescope} \citep[HST, ][]{Bedin2004, Piotto2005, Piotto2007, King2012, Milone2012, Milone2014}, and recently, thanks to the ultra-deep NIR observations with \textit{James Webb Space Telescope} \citep[JWST, ][]{Cadelano2023, Milone2023}.

Although it is known that these peculiar chemical patterns are the footprints of high-temperature H-burning \citep[][and references therein]{Prantzos2017}, the formation scenario of the globular cluster populations and how the processed material finds its way to form the \ac{2P} low-mass stars is still under debate \citep[][for a review]{Bastian2018}.
The main studied formation channels are three self-enrichment scenarios:
the \ac{AGB} scenario, in which stars with initial masses in the range from 6 to 11~\Msun\ are the polluters of the material that form the \ac{2P} \citep{Ventura2001, Ventura2013, DErcole2012, Dantona2016};
the \ac{FRMS} scenario, in which massive rotating stars pollute intracluster gas with the product of the H-burning through the combination of rotational mixing and strong winds \citep{Prantzos2006, Decressin2007b, Decressin2007,  Krause2013};
the \ac{SMS} scenario, where very massive stars ($10^3$ to $10^4~\Msun$) form from runaway collisions of \ac{1P} proto-stars in the core of the proto-cluster.
Such stars forge the processed material and pollute the intracluster medium with processed elements via strong stellar winds \citep{Denissenkov2014, Gieles2018}.
Other interesting scenarios involve the binary evolution of massive stars \citep[e.g., see][]{DeMink2009, Renzini2022} or the formation of \ac{2P} low-mass stars in \ac{1P} massive red super giant stars' shells \citep{Szecsi2018}.
The different globular clusters formation scenarios also help to interpret the unusually high N/O abundance ratios in the distant ($z = 10.6$) star-forming object GN-z11, recently observed with the NIRSpec instrument on board JWST \citep{Cameron2023, Charbonnel2023, DAntona2023, Vink2023}. 

The different formation scenarios predict a different amount of He enrichment of the \ac{2P} stars. 
For instance, in the \ac{AGB} scenario, \citet{Doherty2014} predict an initial helium mass fraction for the \ac{2P} stars of about $0.35-0.4$ due to second dredge-up episodes in the more massive \acp{AGB}.
In the \ac{FRMS} framework, the \ac{2P} stars are predicted to be born with a large variation of initial He (Y) up to 0.8 in mass fraction \citep{Decressin2007}.
These scenarios predict too large a He enrichment with respect to the observed ones, while the \ac{SMS} scenario by \citet{Denissenkov2014} can predict the right amount of enrichment with some fine-tuning to solve the problem of the He overproduction \citep{Bastian2015}.
On the other hand, the He-enrichment can be extremely limited in the case of the SMS conveyor belt scenario by \citet{Gieles2018}. 
Unfortunately, it is difficult to measure the He content of stars directly since most of the stars in globular clusters are too cool to display He lines in their spectra. 
\ac{HB} stars may be hot enough to show weak He lines in their spectra, but they are difficult to model and interpret \citep[also for the role of mixing processes such as atomic diffusion][]{Michaud2008}.
Recently, using the Hubble Space Telescope multi-band photometry, \citet{Milone2018} have indirectly determined the Y abundance of stars in a large homogeneous sample of 57 globular clusters.
They found that the He-enrichment is generally very small and correlates with the clusters' mass, with a maximum He mass fraction of 0.38 in the extreme second population in NGC~2808.
Another indirect constraint comes from the \ac{HB} morphology, which can be explained with a spread in Y \citep[$\delta$Y $\sim 0.02 - 0.15$, e.g., see][]{Tailo2021}. 
\citet{Martins2021} recently redetermined the maximum helium mass fraction of stars in NGC~6752 (computing synthetic populations and spectra of the He-enriched stars) and found that it is unlikely that stars more He-rich than $\sim$ 0.3 are present in the cluster.
It is worth stressing that all the scenarios proposed for candidate polluters, so far, are not able to match all the observational constraints obtained at the same time \citep{Bastian2018}.

\begin{table*}[!ht] 
    \caption{Initial chemical composition, and $\alpha$-enhancement of our stellar tracks for each cluster. We also provide cluster ages from recent investigations to give a reference. Here, we recall that the derived ages depend on the physical parameters used in the stellar codes.} 
    \begin{center}
        \resizebox{\textwidth}{!}{
        \begin{tabular}{rccccccc}
        \hline
        Name  & $\feh_\mathrm{1P}$ & \afe & Z & Y$_\mathrm{1P}$ & Y$_\mathrm{2P}$ & Age$^*$ [Gyr] & Age$^\dagger$ [Gyr] \\ 
        \hline\\
        NGC 4590 & $-2.2$ & +0.35 & 0.0002 & 0.248 & [0.26, 0.27, 0.3, 0.33, 0.37, 0.4, 0.425, 0.45] & 12.03$^{+0.54}_{-0.54}$ & 12.75$\pm 0.75$\\ 
        NGC 6752 & $-1.53$ & +0.3 & 0.0009 & 0.248415 & [0.27, 0.3, 0.4, 0.6] & 13.48$^{+0.81}_{-0.54}$ & 13.00$\pm 0.50$\\ 
        NGC 2808 & $-1.15$ & +0.3 & 0.002 & 0.249 & [0.26, 0.27, 0.3, 0.33, 0.37, 0.40, 0.425] & 10.93$^{+1.20}_{-1.03}$ & 12.00$\pm 0.75$ \\ 
        NGC 6441 & $-0.5$ & +0.2 & 0.008 & 0.255 & [0.27, 0.3, 0.33, 0.37, 0.4, 0.425] & 10.44$^{+2.78}_{-1.62}$ & 12.25$\pm 0.75$ \\ 
        \hline
        \end{tabular}
    }
   	\end{center}
    \footnotesize{$^*$Ages taken from \citet{Valcin2020}. $^\dagger$Ages taken from \citet{Tailo2020}.}
    \label{tab:initial_abund} 
\end{table*}

The initial helium abundance is one of the main ingredients that drive the star's evolution \citep{Iben1968, Iben1969}. The He mass fraction impacts the star's structure and evolution in several aspects.
A higher He content (at fixed Z) leads to a decrease in opacity due to the decrease of the hydrogen mass fraction. 
He-enriched stellar models are generally hotter and brighter than models with standard He in \ac{MS}. 
Consequently, the lifetime of He-enriched stars is significantly reduced, and at a given age, the He-enriched stars are less massive at the \ac{MS} turn-off \citep{Charbonnel2013, Chantereau2015, Cassisi2020}. 
Therefore, if all stars experience the same mass-loss on the \ac{RGB}, those on He-rich isochrones will land on the \ac{HB} in a different position of the \ac{HR} diagram (i.e., with a different effective temperature or color) due to the different mass.

Another interesting effect of a higher initial He content in stellar population is the increased occurrence of so-called ``\ac{AGB}-manqué'' stars \citep[or failed-AGB,][]{Greggio1990, Bressan1993, Dorman1993, Charbonnel2013}.
Among the stars able to ignite and burn helium, some of them end their lives without climbing the \ac{AGB}, and a higher initial He increases the mass range in which such evolution occurs \citep[e.g., see Figure~15 by][]{Chantereau2015}.
Such stars have higher surface temperatures than standard He stars during the \ac{CHeB} phase.
At the end of the \ac{CHeB}, they fail to swell up the transparent and thin He-rich envelope. 
Instead, after the ejection of the external layers, they directly move to the \ac{WD} cooling sequence, avoiding the \ac{AGB} phase.

The \ac{AGB}-manqué scenario helps to interpret some observational properties of globular clusters \citep[see][for a review]{Gratton2019}.
For instance, it was proposed to explain the lack of sodium-rich stars on the \ac{AGB} in NGC~6752 \citep{Charbonnel2013, Cassisi2014}. 
It can also explain the lower fraction of \ac{2P} stars that populate the \ac{AGB} of some clusters revealed in spectroscopic and photometric studies \citep{Campbell2013, Lapenna2016, Marino2017, Lagioia2021}, even though the fraction of second-population AGB stars can be affected by more than one factors \citep{Wang2017}.
Moreover, the \ac{AGB}-manqué prediction is consistent with results by \citet{Gratton2010}, who found that clusters with blue-extended \ac{HB} have a ratio of \ac{AGB} to \ac{RGB} stars smaller than globular clusters without extended \ac{HB}.

In this paper, we follow two objectives. 
The first consists of building new morphologically accurate isochrones from \textsc{starevol} stellar tracks with different initial He content.
The second objective is to study the effects of He-enrichment on isochrones and test them against observed number ratios of stars in selected globular clusters.
This check ensures that our tracks and isochrones are accurate enough to compute integrated synthetic spectra in future dedicated works to study far and not-resolved populations in clusters and galaxies. 
For the computation of this first set of new isochrones, we adopted the sets of tracks by \citet{Charbonnel2016} and \citet{Martins2021}, who used the \ac{FRMS} scenario for the \ac{2P} stars' abundance pattern. 
They computed stellar tracks with four different metallicities (resembling the \feh\ of four GCs) and several initial He contents. 
We took care to extend the few incomplete tracks to the late stages of the \ac{TP-AGB}, and we computed new tracks (when needed) to better resolve the transition between those that undergo \ac{AGB}-manqué and \ac{AGB} phase.
At variance with the work by \citet{Charbonnel2016} and \citet{Martins2021}, in which isochrones are computed up to the \ac{RGB} tip, here we build a consistent set of isochrones, including all the stellar phases from the \ac{ZAMS} to the tip of \ac{TP-AGB}.
To do that, we developed new methodologies to treat all the various types of evolutionary behaviors (typical of low-mass and He-enriched stars) and build morphologically accurate and self-consistent isochrones for the different metallicities, He-enrichments, and ages adopted.
In future works, we plan to compute new isochrones from stellar tracks computed with other formation scenarios and with different adopted physics.

The paper is structured as follows.
In Section~\ref{sec:tracks}, we describe the initial chemical abundances of the adopted stellar models, the input physics, and the stellar tracks' properties.
Section~\ref{sec:Isocs} presents a brief description of the new isochrone building pipeline and the new isochrones (more details are given in Appendix~\ref{sec:app_criteria}).
In Section~\ref{sec:res}, we show and discuss the number ratios obtained for isochrones representing populations of different ages, metallicities, and initial He-enrichment.
Finally, in Section~\ref{sec:concl}, we summarize our results and draw our conclusions.

%--------------------------------------------------------------------
\section{Stellar evolutionary tracks}
\label{sec:tracks}

In this study, we use stellar tracks computed with the \textsc{starevol} code \citep{Siess2000, Palacios2003, Amard2019}. 
We adopted the tracks computed and presented in \citet{Charbonnel2016} and \citet{Martins2021}. 
These stellar tracks follow the evolutions of low-mass stars from the pre-main sequence\footnote{Computations start on the Hayashi track at the beginning of the deuterium burning phase on the PMS that we consider as the time zero of the evolution.} to the \ac{TP-AGB} or planetary nebula phase and \ac{WD} cooling. 
We took care to extend these sets to improve the quality of the isochrones.
In particular, we computed new tracks with masses closer to the transition between different evolutionary patterns, and we completed the unfinished tracks to reach the end of the TP-AGB phase.
The set of tracks spans a mass range from $0.2$ to $1~\Msun$.
In the following, we describe the characteristics of these stellar tracks.

\subsection{Initial abundances for 1P and 2P}
\label{subsec:abund}

The stellar tracks computed by \citet{Charbonnel2016} and \citet{Martins2021} include four values of $\feh = -2.2, -1.53, -1.15$, and $-0.5$, chosen to match those of four globular clusters, namely NGC~4590, NGC~6752, NGC~2808, and NGC~6641. 
For consistency with \citet{Chantereau2017}, the solar-scaled compositions are from \citet{Grevesse1993}. 
For each metallicity, the computed tracks representing the \ac{1P} have an initial helium mass fraction (Y$_{1P}$) that is close to the primordial one, obtained with the relation Y$_{1P}$ = Y$_0$ + ($\Delta$Y/$\Delta$Z) $\times$ Z, where Z is the heavy element mass fraction (metallicity), the Y$_0$ = 0.2463 is the primordial helium mass fraction \citep[from][]{Coc2013}, and $\Delta$Y/$\Delta$Z = 1.62 is derived from the solar calibration \citep[by ][]{Grevesse1993} and primordial helium abundances. 
For each \feh, they adopt an $\alpha$-enhancement following \citet{Carretta2010}, as listed in Table~\ref{tab:initial_abund}. 
We stress that, at variance with the previous papers of this series \citep{Chantereau2017}, to better match the metallicity of NGC~6752, we use the tracks by \citet{Martins2021}, who assume a metallicity of $\feh = -1.53$ instead of $\feh = -1.75$.

For the \ac{2P}, \citet{Charbonnel2016} and \citet{Martins2021} adopted the He-Na correlation predicted by the \ac{FRMS} scenario,  described also in \citet{Charbonnel2013}.
In particular, \ac{2P} stellar models are initially depleted in C, O, Mg, Li, Be, and B, and enriched in He, N, Na, and Al to various degrees \citep{Chantereau2016}. 
Stellar tracks representing the \ac{2P} are computed with several initial helium abundances ranging from $\sim 0.26$ to $0.6$ (in mass fraction), assuming the \ac{FRMS} scenario.
The C+N+O content is kept constant between the two populations since the \ac{FRMS} scenario predicts that He-burning products are not included in \ac{2P} stars in agreement with the observations. 
These assumptions are compatible with the predictions of the conveyor-belt \ac{SMS} scenario, modulo the fact that high Na and Al enrichment are reached already for much lower He-enrichment because of the highest H-burning temperature reached within these objects.
Table~\ref{tab:initial_abund} lists all the different metallicities $\feh$ and initial helium abundances used to compute the stellar tracks we adopted in this work. 

\subsection{Adopted physics}
\label{subsec:phys}

The basic input physics adopted for the stellar models can be summarized as follows:
\begin{itemize}
    \item We follow stellar nucleosynthesis of 54 chemical species from $^1$H to $^{37}$Cl, with a network of 185 nuclear reactions \citep[see][for more details]{Lagarde2012}. 
    We use the nuclear reaction rates from the NACRE2 database generated using the NetGen web interface \citep{Xu2013a, Xu2013b}. 
    \item The screening factors are included using the \citet{Mitler1977} and \citet{Graboske1973} formalism. 
    \item Opacities are interpolated from the OPAL\footnote{\url{https://opalopacity.llnl.gov}} project opacity tables \citep{Iglesias1996} for $T > 8000$ K, and atomic and molecular opacity by \citet{Ferguson2005} for $T < 8000$ K.
    \item For the equation of state, we adopt the formalism by \citet{Eggleton1973} and \citet{Pols1995} \citep[see][for more details]{Siess2000}.
    \item Convection is treated within the mixing-length theory \citep{Bohm-Vitense1958}, assuming a parameter $\alpha_\mathrm{MLT} = 1.75$ \citep[as adopted in the \ac{FRMS} scenario by ][]{Decressin2007}.
    We assume instantaneous mixing in the convective regions, except during the \ac{TP-AGB} phase, in which we adopt a time-dependent diffusion treatment to follow the hot-bottom burning properly \citep{Forestini1997}.
    \item To determine the regions unstable to convection, we use the \citet{Schwarzschild1958} criterion.
    \item We use the gray atmosphere approximation for the photosphere, defined as the layer with an optical depth $\tau$ between 0.005 and 10. The effective temperature, \Teff\, and the stellar radius are defined at the layer where $\tau = 2/3$.
    \item Mass loss is taken into account, assuming the \citet{Reimers1975} prescription (with $\eta = 0.5$) on the \ac{RGB} and during \ac{CHeB}. During the \ac{AGB} phase, we use the \citet{Vassiliadis1993} recipe for the oxygen-rich stars and switch to \citet{Arndt1997} whenever the low-mass AGB becomes C-rich. 
    The mass loss in such phases (particularly on the \ac{RGB}) is still an open issue and has great uncertainty.
    The common prescription adopted by stellar modelers for this phase is the \citet{Reimers1975}, with a parameter $\eta$ that can vary from 0.1 to 0.65, depending on the calibration performed \citep[e.g., see][]{ Miglio2012, McDonald2015, Tailo2021}.
    In the choice of this parameter, \citet{Charbonnel2016} and \citet{Martins2021} used the results by \citet{McDonald2015}, who give a median determined from 56 GCs of $\eta = 0.477\pm0.07$.
\end{itemize}
In the models, the TP-AGB phase is fully followed within the \textsc{starevol} stellar evolution code \citep{Forestini1997}.
This means that contrary to models computed by synthetic codes \citep{Groenewegen1993} and hybrid `envelope-based' codes such as the \textsc{colibri} code \cite{Marigo2013}, the main ingredients impacting the TP-AGB evolution, e.g. mass loss, efficiency of mixing, depth of the third dredge-up, core growth, cannot be fine-tuned and are fully dictated by the physical prescriptions adopted from the beginning of the stellar evolution. 
In the present models, $s-$process element nucleosynthesis is not followed.

The stellar tracks are computed without atomic diffusion, semi-convection, thermohaline mixing, rotation, and overshooting.
Neglecting atomic diffusion and rotation induces an uncertainty on the low-mass stars' lifetime of about 10 to 20$\%$ \citep[e.g.][ see Appendix~\ref{sec:app_comp} for a comparison]{VandenBerg2002, Lagarde2012, Dotter2017, Borisov2024}, which can be of the same order of errors obtained from \ac{GC} age estimations \citep[e.g., for NGC~6752 13.48 $\pm$ 0.7 Gyr,][]{Valcin2020}.
The \ac{FRMS} scenario predicts a maximum delay of tens of Myr between the formation of the two stellar populations \citep{Krause2013}. 
The observational constraints coming from young, very massive, compact, and gas-free star clusters point rather to less than 10~Myr \citep{Krause2016}, which is compatible with the SMS scenario and anyway negligible compared to the \ac{GC}'s age uncertainty.
In future work, we plan to use sets of stellar tracks based on different formation scenarios than the \ac{FRMS}, including also the mentioned physical processes, to study their impact on the \ac{2P} stars. 

\begin{figure}
    \centering
    \includegraphics[width=\columnwidth]{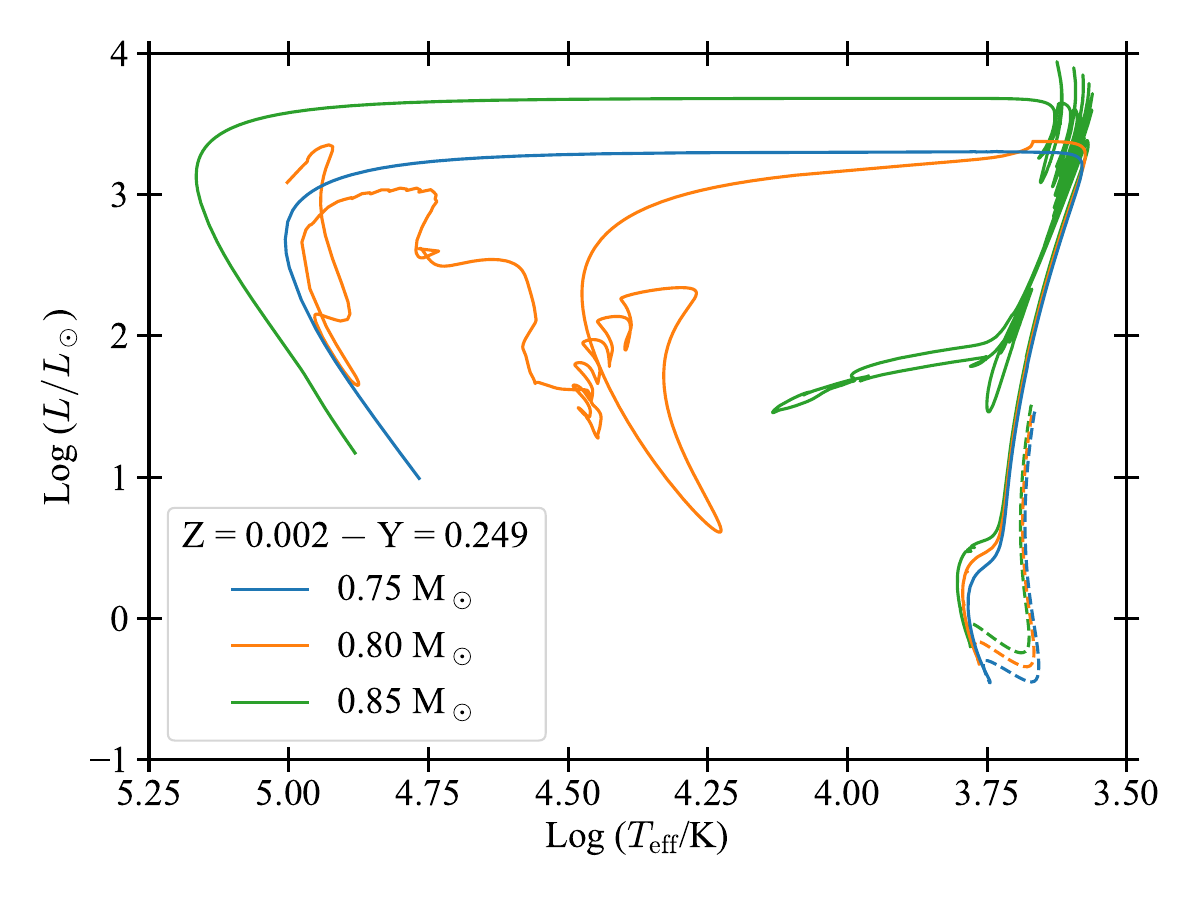}
    \caption{Hertzsprung-Russell (HR) diagram of three selected stellar evolutionary tracks computed with \textsc{starevol}, with different evolutionary paths. The dashed lines show the pre-MS phase.}
    \label{fig:HR_stellar_paths}
\end{figure}

\subsection{Stellar evolutionary paths}
\label{subsec:paths}

\begin{figure*}
   \centering
        \includegraphics[width=\columnwidth]{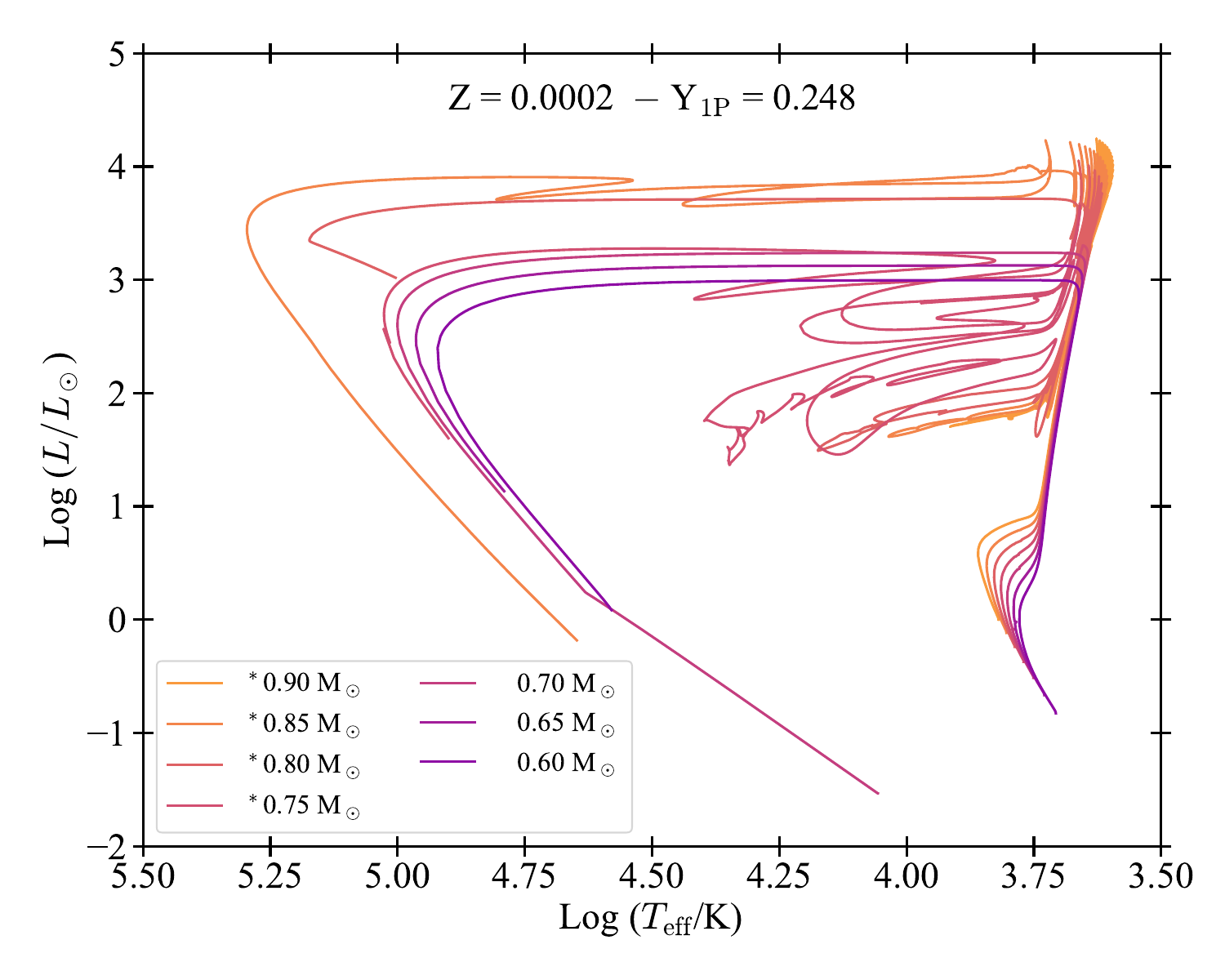}
        \includegraphics[width=\columnwidth]{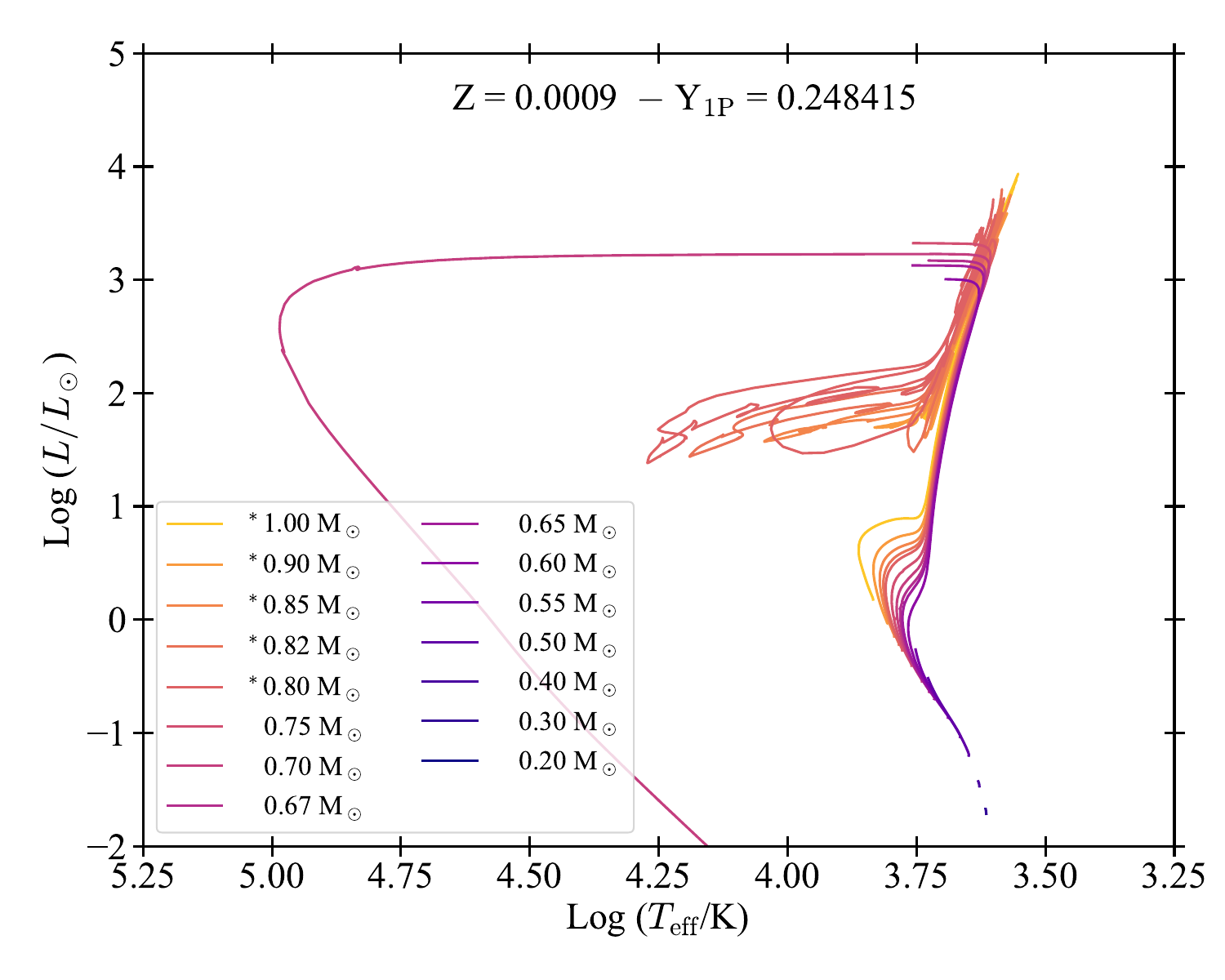} \\
        \includegraphics[width=\columnwidth]{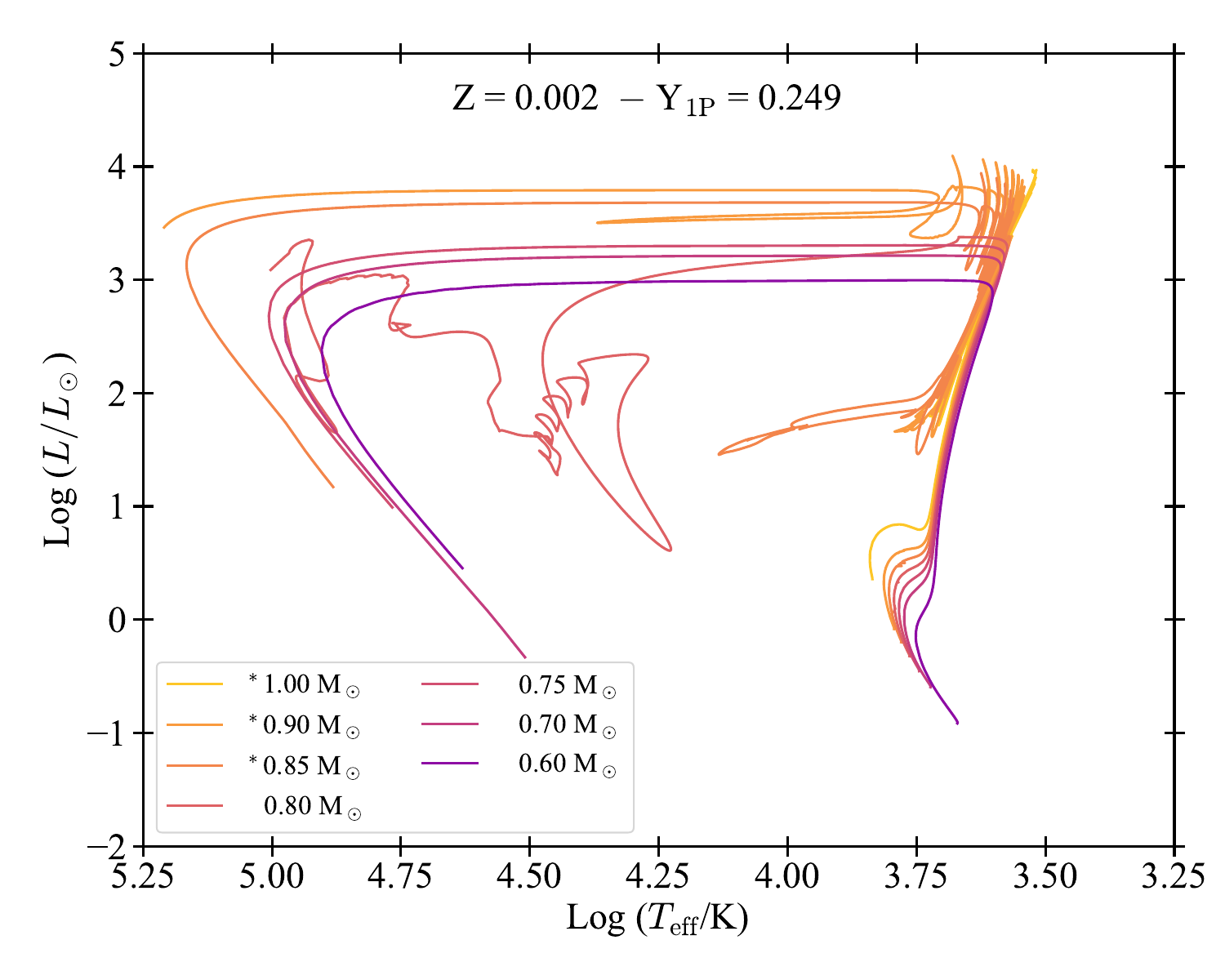}
        \includegraphics[width=\columnwidth]{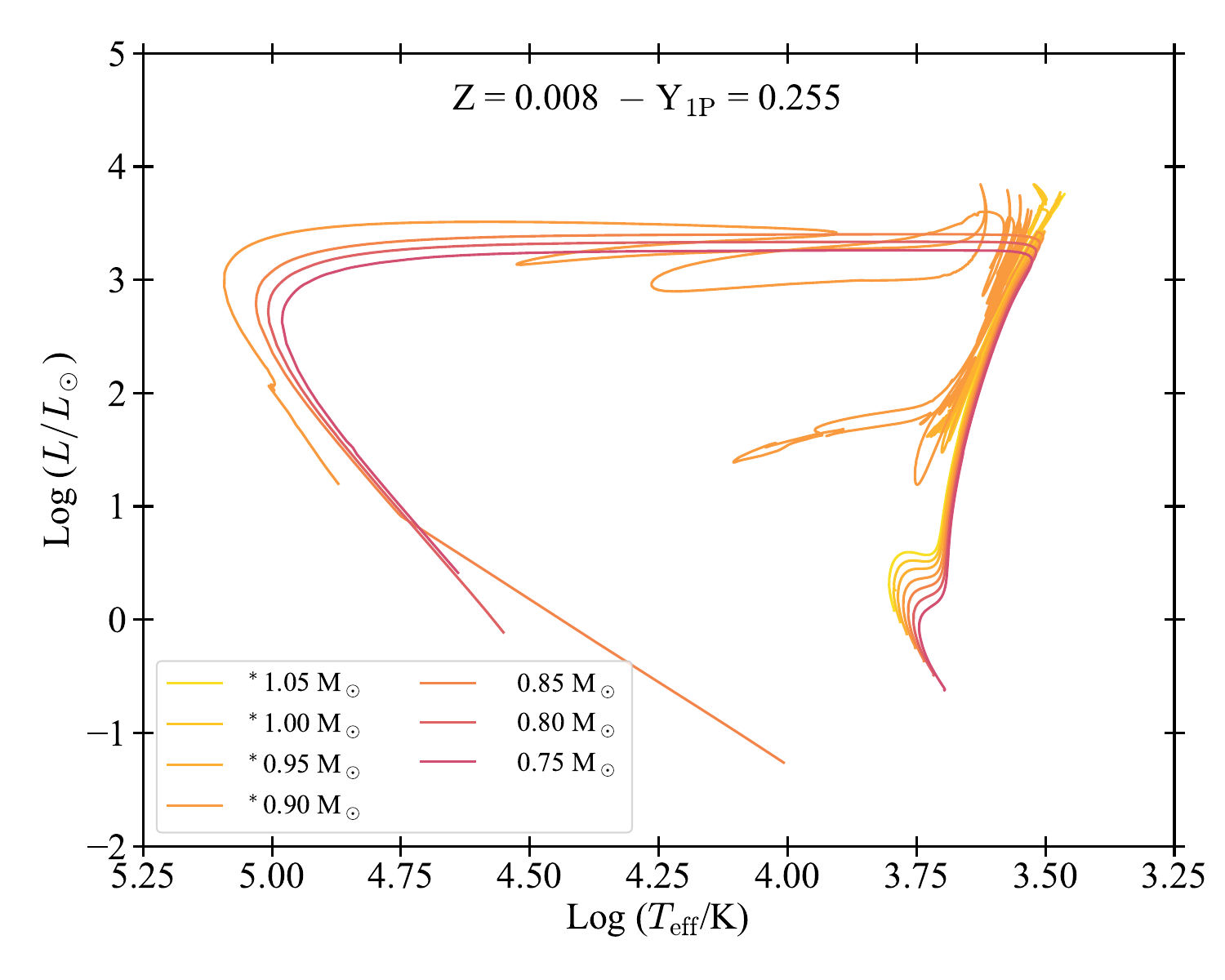}
   \caption{Selected evolutionary tracks of \textsc{starevol} in the HR diagram, representing \ac{1P} stars of the four cluster listed in Table~\ref{tab:initial_abund}. Starting from the top-left and going clockwise, tracks are computed with Z = 0.0002, 0.0009, 0.008, and 0.002. Different colors indicate different initial masses. Tracks are shown from their \ac{ZAMS} to the planetary nebula phase and \ac{WD} cooling. The little star apex in the label before the masses indicates stars that go to the \ac{AGB} phase.}
   \label{fig:HR_1P}
\end{figure*}

To illustrate the various evolutionary paths that low-mass stars may follow, we show a set of tracks with Z = 0.002 and Y = 0.249 in Figure~\ref{fig:HR_stellar_paths}. 
The selected tracks show the three different evolutionary paths.
For this initial composition, we find that tracks with an initial mass $M_\mathrm{ZAMS} \leq 0.75~\Msun$ are not massive enough to ignite helium.
Near the \ac{RGB} tip, they experience strong stellar winds that peel off their envelope, exposing the degenerate helium core. 
When the envelope is almost totally expelled, the stars move to the planetary nebula phase and then to the \ac{WD} cooling stage. 

Stars with $0.75 < M_\mathrm{ZAMS}/\Msun < 0.85$ are massive enough to ignite helium at the \ac{RGB} tip, undergoing the so-called He-flash. These stars then move to the \ac{HB} where they undergo stable \ac{CHeB}.
In the fast transition phase, which lasts for $\sim 1.3$ Myr, they may experience several secondary He-flashes (less luminous than the first one) that manifest as loops in the \ac{HR} diagram.
Stars in this mass range retain only tiny envelopes, and they lie on the blue side of the \ac{HB} during the \ac{CHeB}. 
They have higher effective temperatures than more massive stars that retain more massive envelopes \citep[see, e.g.][]{Rood1973}. 
The mass of the envelope also affects the post-\ac{CHeB} phase because it is too tiny to allow the shell-burning processes that - in the usual case - force stars to reach the \ac{AGB} stage.
Therefore, these stars with tiny envelopes may completely miss the \ac{AGB} phase.
Moreover, the stellar winds could eject the small envelope during the \ac{CHeB} phase. 
Therefore, the star cannot expand and completely avoids the evolution toward the \ac{AGB}, moving instead to higher luminosities and directly to the planetary nebula phase and \ac{WD} cooling sequence. 
This is the evolutionary pattern of \ac{AGB}-manqué stars.

For stars with a higher initial mass that also have higher envelope mass when they reach the RGB tip, the post-\ac{CHeB} evolution follows the `standard' behavior.
During the \ac{CHeB}, tracks with $M_\mathrm{ZAMS} \geq 0.85~\Msun$, have a lower effective temperature, and after this phase, they move to the \ac{AGB} phase. 
The advanced \ac{AGB} phase is characterized by thermal pulses, dredge-up, and strong stellar winds, which determine the duration of this phase.
When stellar winds almost totally peel off the envelope, stars move to the planetary nebula phase (sometimes experiencing a last thermal pulse) and finally to the \ac{WD} cooling sequence.

Figure~\ref{fig:HR_1P} shows all sets of stellar tracks we use in this work to represent stars of the \ac{1P} of the four clusters. 
After the \ac{MS}, all stars cross the sub-giant region and climb the \ac{RGB}.
Subsequently, as described above, stars may follow three different evolutionary paths, depending on the stellar structure at the tip of the \ac{RGB}.
We consider each track complete if the star has an envelope < 0.1~\Msun\ during the planetary nebula phase.
In this assumption, all our tracks are complete for our purposes (also those with Z = 0.0009 and Y = 0.248415 shown in the top-right panel of Figure~\ref{fig:HR_1P}).

The transition masses between the different kinds of evolution and the occurrence of \ac{AGB}-manqué behavior depend on the initial abundances of helium and metals \citep[as shown in figures 5 and 6 by][]{Charbonnel2016}. 
In general, the higher the metal content, the higher the transition mass between stars with the \ac{AGB}-manqué and \ac{AGB} evolutionary patterns. 
On the other hand, for \ac{2P} stars, the trend of the transition masses is not linear with the initial helium abundance. 
For Y < 0.5, the higher the helium, the smaller the transition mass.
While, in cases of a very high helium enrichment (i.e., Y > 0.5), the trend can be inverted \citep[see Figure 2 in][]{Chantereau2016}.
From the analysis presented in Appendix~\ref{sec:app_comp}, we expect the effect of detailed abundances of elements other than He and which do count in the metal mass fraction Z to be small.
We thus expect that the use of 2P abundances derived from the FRMS scenario, as done here or from other scenarios, will not impact the results obtained in this work.
For a detailed description of the impact of the helium enrichment on the tracks and the transition masses for different metallicities, we refer the reader to \citet{Chantereau2015} and \citet{Charbonnel2016}.

It is worth noticing that the transition masses and the occurrence of the \ac{AGB}-manqué evolution also depend on the mass-loss in the \ac{RGB} and \ac{HB} phases.
The mass loss in such phases (particularly on the \ac{RGB}) is still an open issue and has great uncertainty \citep[e.g.][]{Tailo2021}.
Further exploration of the uncertainty related to stellar wind prescriptions is outside the scope of this paper.

\section{Isochrones}
% \section{New isochrones}
\label{sec:Isocs}

\begin{figure}
    \centering
    \includegraphics[width=\columnwidth]{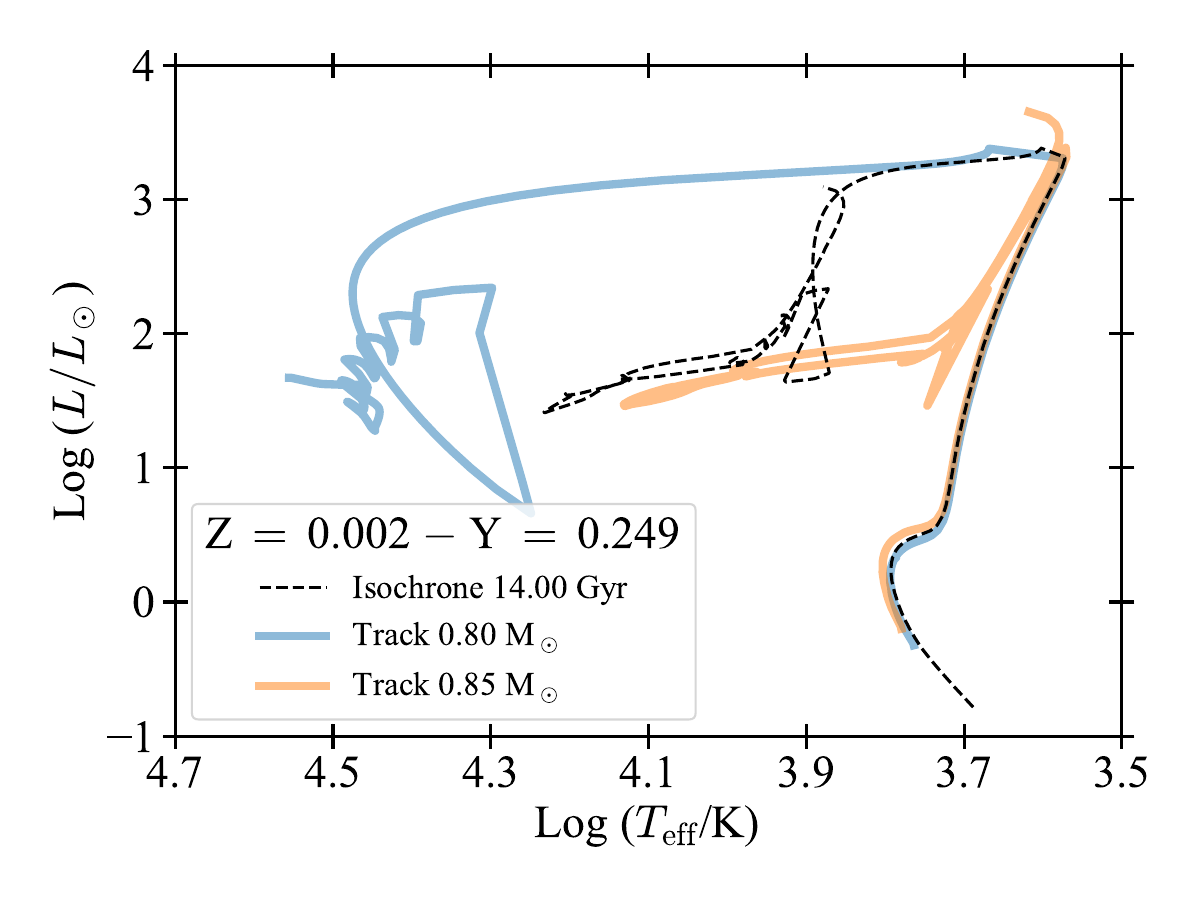}
    \caption{Example of an incorrect isochrone (dashed black line), as obtained when omitting to account for the different morphologies of the post-\ac{RGB} phase of the two tracks used in the interpolation.
    The isochrone has an age of 14 Gyr. 
    The two tracks used for the interpolation of the post-\ac{RGB} phases are shown in blue and orange. The initial metallicity is Z=0.002, and the initial He mass fraction is 0.249. 
    The two tracks shown are reduced tracks obtained following the method described in Section~\ref{subsec:cp}.
    }
    \label{fig:HR_without_fakestar}
\end{figure}

The computation of theoretical isochrones has a long tradition, and several groups have developed their tools to tackle this task \citep[some recent examples are][]{Vandenberg2014, Georgy2014, Dotter2016, Spada2017, Hidalgo2018, Nguyen2022}.
These tools use smart and efficient methods that exploit the similarities between tracks to obtain accurate isochrones, starting from sets of tracks, which, for obvious reasons, cannot span the space of initial parameters `continuously'.
One common key idea adopted in the isochrones building methods is the selection of the so-called \textit{equivalent evolutionary points} (or critical points) along the evolutionary tracks  \citep{Simpson1970, Bertelli1990, Bertelli1994, Girardi2000, Ekstrom2012}. 
An accurate selection of these critical points permits obtaining smooth interpolated tracks and morphologically realistic isochrones without wasting resources in the computation of very big stellar track grids with, e.g., infinitely small mass steps between each track.
Typically, grids of tracks to compute isochrones include a few hundred tracks, not more.

It is worth stressing here that a good selection of critical points is not enough to obtain acceptable isochrones. Since stellar tracks with two very close masses may show a very different morphology in the post-\ac{RGB} phase (as shown in Fig.~\ref{fig:HR_stellar_paths}), a specific strategy to treat this aspect accurately must be adopted. Ignoring those transitions in the isochrone building methods will produce wrong isochrones, as the one shown in Fig.~\ref{fig:HR_without_fakestar}.
In this case, the isochrone in the post-\ac{RGB} is computed from the interpolations of the two tracks shown without an adapted methodology. This isochrone is a clear example of a `blind' interpolation that produces nonphysical properties, such as a faint ($\log L < 3$) and hot ($\Teff > 3.8$) \ac{AGB} phase. Because the stars in these phases are bright, their properties impact the integrated colors of a synthetic stellar population. Some of the evolutionary phases affected are short-lived, in which case their impact on average integrated properties is limited, but they nevertheless modify the statistical distributions of colors of equal-age populations of synthetic clusters with finite numbers of stars (a distribution due to the stochastic sampling of the IMF; for examples see \citealt{GirardiBica1993, Fouesneau2010}). 

\subsection{Isochrone building with \textsc{syclist}} 
\label{subsec:SYCL}

Our building methodology consists of three steps: the creation of reduced stellar track tables, the creation of the so-called `fake' tracks, and the isochrone-building process.
All the details are described in Appendix~\ref{sec:app_criteria}.
In brief, in the first step, we select the so-called equivalent evolutionary points and interpolate between them to create reduced tables. In the second, we create the `fake' tracks needed to avoid the inadequate interpolations between tracks with different post-\ac{RGB} morphology.
These tracks share the evolutionary properties of the two nearest tracks that enclose the transition.
For the final step that leads from properly re-sampled tracks to isochrones, we used the \textsc{syclist} code \citep[described in detail in][]{Georgy2014}.
Thanks to its flexible design, modifying the equivalent points along tracks and introducing the `fake' track method did not change the code's structure.
We updated the code to propagate several surface abundances from the \textsc{starevol} stellar tracks to the isochrones, which will be useful for any subsequent computation of detailed synthetic spectra.
Our isochrones now include surface abundances of 40 isotopes from $^1$H to $^{37}$Cl, plus all the remaining heavier elements collected in a single variable, `Heavy'.
The heavier isotopes (included in the `Heavy' abundance) could be easily retrieved using the initial element partition adopted since these elements do not evolve in time in our stellar models.

\begin{figure}
    \centering
    \includegraphics[width=\columnwidth]{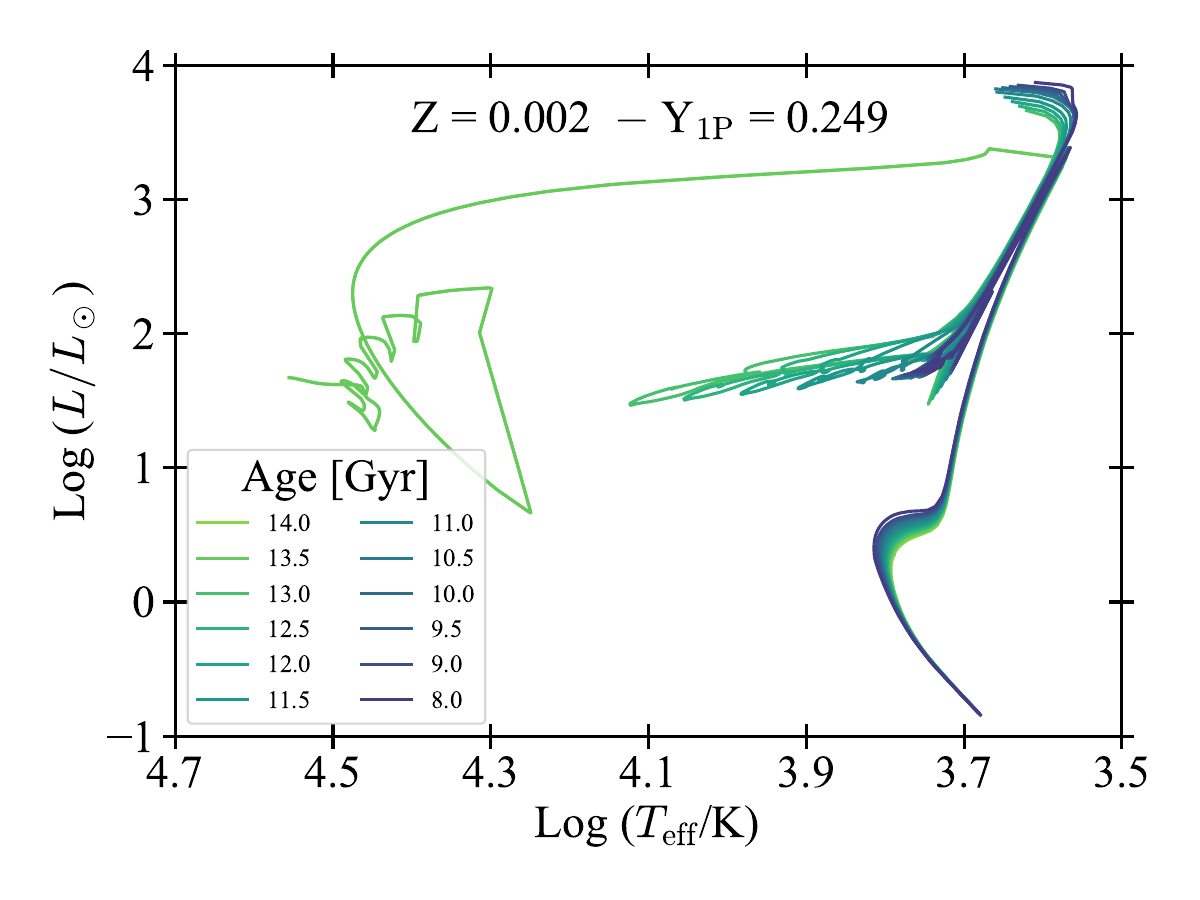}
    \caption{The figure shows isochrones with different ages from 8 to 16 Gyr with different colors. The initial abundances are Z = 0.002 and Y = 0.249.
    Isochrones with 13.5 and 14 Gyr overlap due to our isochrone-building methodology. Therefore, only the one with 13.5 Gyr is visible.
    }
    \label{fig:Isocs}
\end{figure}

\begin{figure}
    \centering
    \includegraphics[width=\columnwidth]{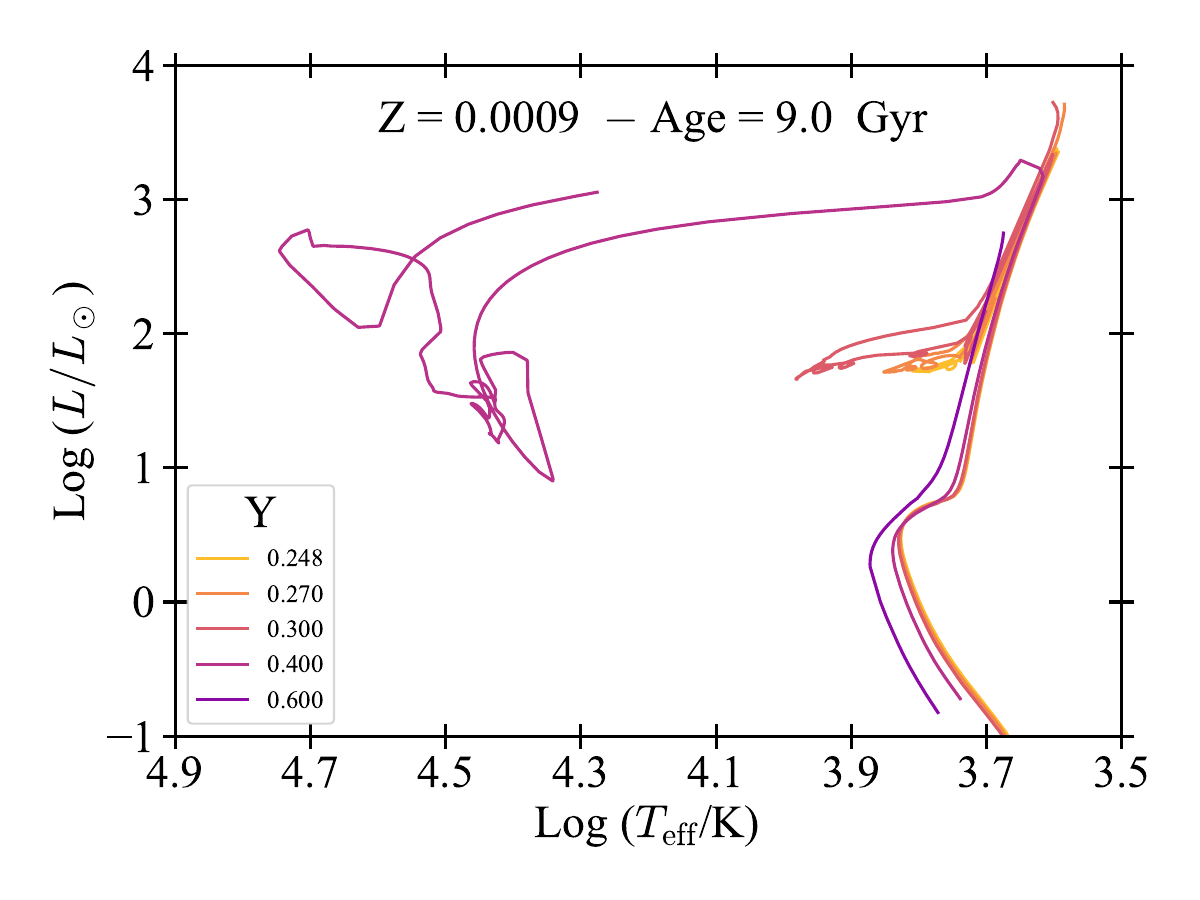} \\
    \includegraphics[width=\columnwidth]{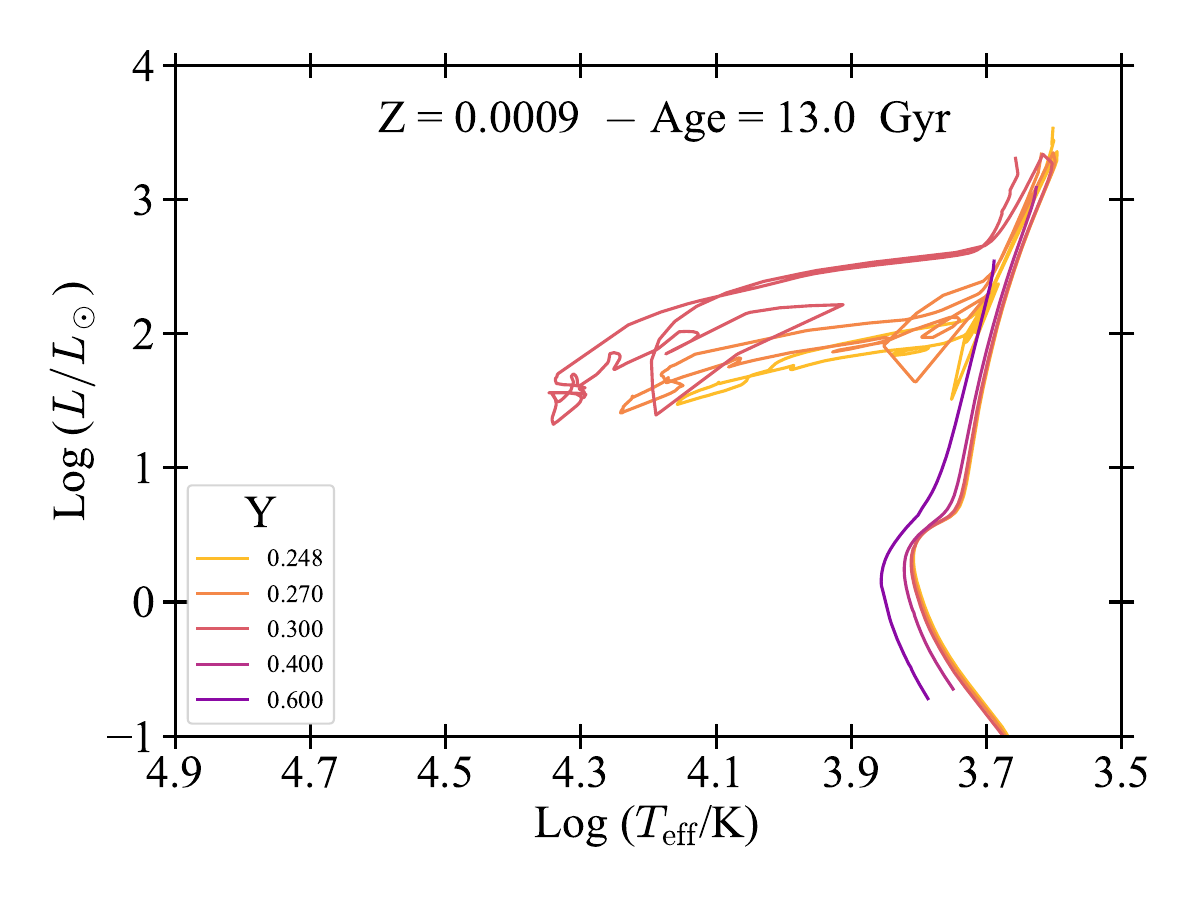}
    \caption{The top and bottom panels show isochrones with ages 9~Gyr and 13~Gyr, respectively. Different colors indicate the different initial He content. All isochrones shown here have Z = 0.0009.
    }
    \label{fig:Isocs_Y}
\end{figure}

\subsection{New isochrones}
\label{subsec:res_isoc}

For each set listed in Table~\ref{tab:initial_abund}, we computed new isochrones from 9~Gyr to 14~Gyr, with steps of 0.5~Gyr.
Figure~\ref{fig:Isocs} shows an example of selected isochrones computed with a fixed initial composition and different ages.
The figure shows how, as the age of the isochrones increases, the turn-off becomes fainter and redder, and the star's position in the \ac{HB} becomes bluer.
The bluer position of the \ac{HB} at older ages is related to the smaller stellar mass at the \ac{RGB} tip and the smaller envelope masses that such stars retain in that phase \citep[][]{Bertelli2008}.
When the envelope mass is very small, the \ac{AGB}-manqué evolutionary pattern appears. 
In this case, stars burn He in a very hot location on the \ac{HR} diagram and then evolve directly to the planetary nebula, skipping the \ac{AGB} phase.
In Figure~\ref{fig:Isocs}, all isochrones aged over 13~Gyr show the \ac{AGB}-manqué behavior (note that these isochrones overlap in the figure because of our isochrone building methodology).
While isochrones with ages $\leq 13$~Gyr have a colder \ac{HB} phase and stars in the \ac{AGB} phase.

Figure~\ref{fig:Isocs_Y} shows selected isochrones with different initial He and two different ages but the same metallicity. 
Isochrones with different initial He-content run in the \ac{HR} diagram almost parallel during the \ac{MS}.
As already mentioned, a higher initial He takes to a lower mean opacity and a higher mean molecular weight in the stellar envelope, leading to hotter effective temperatures.
Counterintuitively, He-enriched populations are fainter in the \ac{MS} turn-off because of the shorter \ac{MS} lifetimes, which indicate a lower mass at the turn-off \citep{Salaris2005, Bertelli2008, Chantereau2015, Cassisi2020}.
After the \ac{MS}, the initial He-content does not affect the morphology of the sub-giant branch much. 
Still, He-enriched populations climb the \ac{RGB} with higher effective temperatures.
The He-content also affects the luminosity tip of the \ac{RGB}.
With an increasing initial He abundance, the brightness of the tip decreases.
This is due to the higher central temperatures of the He-enriched models, which take to lower degenerate cores and to a faster rate of growth of the core.
Both these effects make the He-flash condition occur with smaller He core masses, hence with lower luminosities \citep{Cassisi2020}.
After the He-flash, the initial He content affects the isochrones' morphologies for two reasons.
The first is that the He-enriched populations reach the \ac{RGB} tip with a smaller ratio between the envelope and total mass, mainly due to the smaller \ac{MS} turn-off masses.
The second is that the abundance of He in the stellar envelope during the \ac{HB} is higher in He-rich stars.
They both take the He-enriched populations to have hotter (i.e., bluer) \ac{HB} than populations with standard He.
For the same reasons above, the occurrence of \ac{AGB}-manqué in a stellar population at a fixed age also depends on the initial He.
In Figure~\ref{fig:Isocs_Y}, isochrones with 9 and 13~Gyr, and with Y < 0.4 have models in the \ac{AGB} phase, and those with Y = 0.4 have \ac{AGB}-manqué stars.
Finally, models in isochrones with Y = 0.6 just reach the \ac{RGB}-tip before going to the planetary phase and \ac{WD} cooling sequence.

\section{Number ratios}
\label{sec:res}

The frequency of \acp{AGB} with respect to \acp{RGB} and \acp{HB} stars is strictly related to multiple populations in globular clusters \citep{Gratton2019}. 
Since \ac{AGB}-manqué stars at certain ages are associated with the He-enriched populations of a cluster \citep{DAntona2002, Charbonnel2013, Chantereau2016}, differences in the initial He-enrichment affect the ratios $R_1 = N_{AGB}/N_{RGB}$ and $R_2 = N_{AGB}/N_{HB}$ (which are proportional to the lifetimes' ratios of these phases). 
Our tracks include all the evolution from the \ac{ZAMS} to the \ac{TP-AGB}; thus, we can consistently compute and predict values of $R_1$ and $R_2$ for all our isochrones with different ages and initial compositions and compare them with those from the literature obtained from the observed data.
The purpose of this comparison is to check the accuracy of our new isochrones, which will be used to compute integrated synthetic spectra in a dedicated work to study distant and not-resolved populations.

\begin{figure*}
   \centering
        \includegraphics[width=\textwidth]{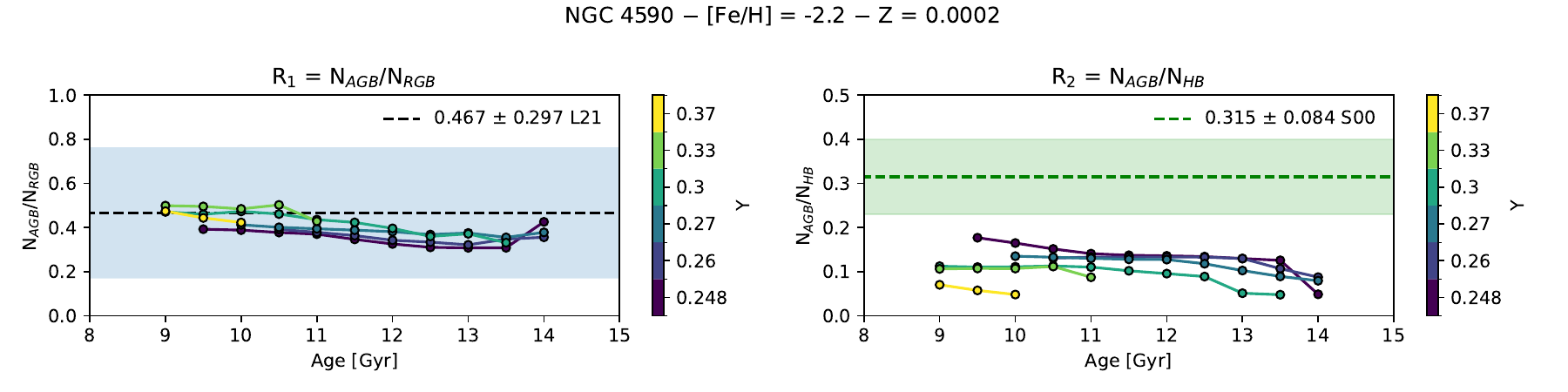}\\
        \includegraphics[width=\textwidth]{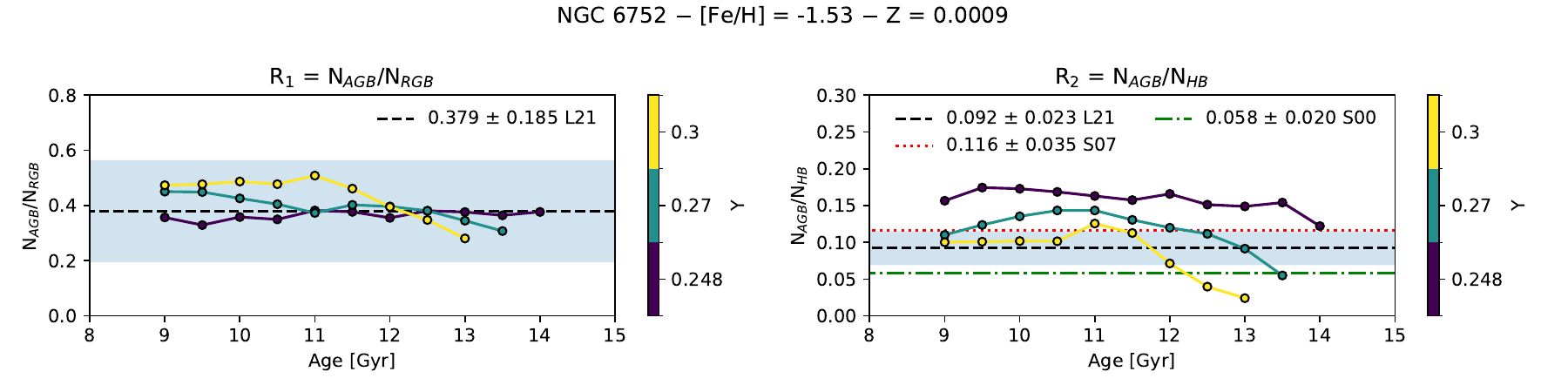}\\
        \includegraphics[width=\textwidth]{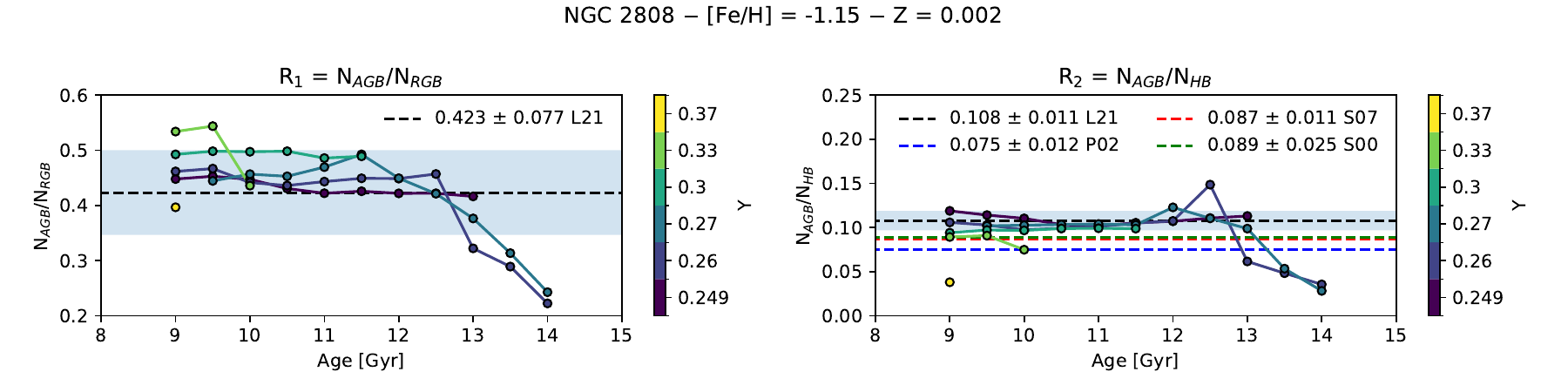} \\
        \includegraphics[width=\textwidth]{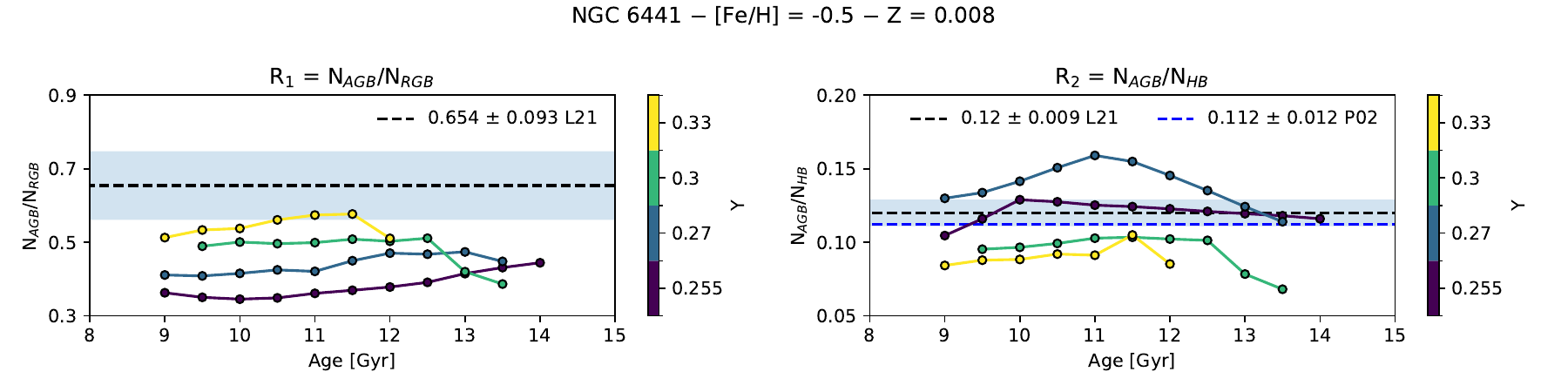}
   \caption{Number ratios versus age, for different initial He content (in different colors). The left column shows $R_1$, while the right column shows $R_2$. Each row indicates the ratios for different initial metallicity (i.e., for different clusters). 
   % The error bars indicate 1$\sigma$ statistical Poisson error.
   The dashed horizontal black lines indicate the number ratios for each cluster obtained by \citet[][L21]{Lagioia2021} from observations. 
   The corresponding 1$\sigma$ statistical Poisson error is indicated by the blue horizontal area.
   The dashed blue and red lines indicate $R_2$ results by \citet{Constantino2016}, which used data from \citet[][P02]{Piotto2002} and \citet[][S07]{Sarajedini2007}, respectively. 
   The dashed green line shows values from the collection by \citet[][S00]{Sandquist2000}.
   The green area in the upper-right panel shows the error associated with the \citet{Sandquist2000} value of the ratio.
   For clarity purposes, we do not show the statistical errors of all observed values in the other panels, but only those by \citet{Lagioia2021}. 
   The other errors have amplitudes similar to those shown here.}
   \label{fig:Ratios}%
\end{figure*}

\subsection{Ratios computation}
\label{subsec:ratios}

Assuming a certain age, the number of stars, $N_X$, of a population in an evolutionary phase, $X$, could be computed as follows
\begin{equation}
    N_X = \int_{\Delta M_X} \Phi(M_{ZAMS})\, \delta M_{ZAMS},
\end{equation}
where $\Delta M_X$ is the interval of mass within the selected phase $X$, $\Phi(M_{ZAMS})$ is the initial mass function and $M_{ZAMS}$ is the initial stellar mass.
By knowing the mass limits of the different phases from our isochrones and assuming a Kroupa initial mass function \citep{Kroupa2001}, we can compute the number ratios $R_1$ and $R_2$ for each isochrone of our sets.
We define all the stellar phases in theoretical isochrones by checking the internal properties of the models (i.e., using the same approach described in Sec.~\ref{subsec:cp}).
On the other hand, a precise definition of the stellar phases cannot be easily retrieved from observations. 
In particular, it is difficult to disentangle stars in the \ac{RGB} and \ac{AGB} phases, which are often quite overlapped in optical CMDs due to their similar effective temperatures. 
Therefore, some strategy must be adopted to decouple such stars and break the degeneracy.
An interesting methodology has been used by \citet{Lagioia2021}, who analyzed photometric data of several globular clusters from HST. 
They used far-UV-optical CMDs to break the color degeneracy of \ac{RGB} and \ac{AGB} stars, exploiting the property of \ac{AGB} stars that are significantly brighter than \ac{RGB} stars in the UV flux.
After the selection of the \ac{AGB} stars, they selected the fainter \ac{AGB} star in optical CMDs (F814W band) and used that magnitude as a limit for selecting the brighter \ac{RGB} stars.
This methodology ensures a reasonable comparison between stars in the two phases.
To compare our theoretical ratios with the empirical ones derived from observational data, we adopt a strategy inspired by \citet{Lagioia2021} to derive stellar numbers of \ac{RGB} stars. 
This step is required to compare our $R_1$ number ratios with theirs.
Firstly, we selected the luminosity of the early-\ac{AGB} phase (defined as the onset of the He-shell burning). Then, we used that luminosity as a lower limit to select the mass limit of the most luminous models in the \ac{RGB} phase.
As regards the \ac{HB} phase, the selected mass limits are the beginning of the \ac{CHeB} and the beginning of the early-\ac{AGB} phases.
Once the mass limits of each phase are defined, we computed the number ratios $R_1$ and $R_2$ for all the isochrones in our sets.
We compute the ratios only for isochrones that populate the \ac{AGB}, i.e., with models that lie in the region of the \ac{HR} diagram defined by $\Teff < 3.85$ and $\log L > 3.1$, making possible the computation of the ratios.
We also associate a statistical Poisson error with the observed ratios when the authors do not provide one.
Since the error size depends on the cluster's number of \ac{AGB} stars, number ratios of clusters with a few \acp{AGB} (such as for Z = 0.0002 and 0.0009 that correspond to NGC~4590 and NGC~6752) have bigger errors.
On the modeling side, it is important to say that there are many uncertainties on several physical stellar processes that can change the number counts, such as the mass loss in the \ac{RGB} and \ac{AGB} phases, the thermohaline mixing, the semiconvection, stellar rotation, and nuclear rates uncertainties during the \ac{CHeB} phase.
In Appendix~\ref{sec:app_comp}, we show a partial comparison with stellar tracks found in the literature and using the \textsc{starevol} code with different physical assumptions.

\subsection{Results and comparison with data}
\label{subsec:comp_data}

The ratios for each initial composition and age are shown in Figure~\ref{fig:Ratios} and listed in Table~\ref{tab:ratios}.
Our results indicate a general trend for all cases. 
We find that $R_1$ is proportional to the initial He-enrichment, while $R_2$ has an inverse proportionality with the initial He. 
This is expected since the lifetimes of such phases depend on the initial He and because $(d \tau_{RGB}/ d Y_{ini}) < (d \tau_{AGB}/ d Y_{ini}) < (d \tau_{HB}/ d Y_{ini}) $, where $\tau_{RGB}$, $\tau_{AGB}$ and $\tau_{HB}$ are the lifetimes of the different phases. 
An example of the lifetime trends with the initial He is shown by \citet{Chantereau2015} in Figure~3.
However, this general trend may change for $R_1$ at older ages ($12 - 13$ Gyrs) due to the \ac{AGB} lifetimes of the He-enriched populations.
In particular, stars close to the AGBm-AGB transition mass ($\delta M$ < 0.05~\Msun) have shorter \ac{AGB} lifetimes than stars with a mass well above the transition ($\delta M$ < 0.05~\Msun).

We also compare our number ratios with values from the literature obtained from observations.
In particular, we compare the $R_1$ ratios with those obtained from observation by \citet{Lagioia2021} because they are retrieved with similar methodologies (as described above). 
Regarding $R_2$, we compare the ratios with those by \citet{Sandquist2000}, \citet{Constantino2016} who computed the ratios from different catalogs \citep{Piotto2002, Sarajedini2007}, and by \citet{Lagioia2021}.
The observed values are retrieved by counting all stars in the cluster, thus, without distinguishing between 1P and 2P stars.
The comparison is not intended as an attempt to fit the four clusters in detail, but only as a test for our post-RGB tracks and for the new isochrones, which are representative of a simple stellar population (i.e., a coeval population with a certain initial abundance). 
This check has the aim of verifying whether the current set of evolutionary models, without any fine-tuning, reproduces lifetimes in the various post-RGB phases reasonably well.

From the comparison of the number ratios with the observed ones, we find that our $R_1$ and $R_2$ roughly agree within 1$\sigma$ with observations. 
There are only two exceptions, one for the $R_1$ ratios computed with Z = 0.008 and the other for the $R_2$ ratios for Z = 0.0002. 
In the first case, the theoretical ratios are consistently lower than the observed value by 1 or more $\sigma$, depending on the initial He.
In this case, He-enriched populations seem to reproduce better the observations.
In the second case, the $R_2$ values are off the observed value by more than 1$\sigma$.
However, it is worth noticing that the $R_2$ value presented by \citet{Sandquist2000} for this cluster should be taken with care since the number of \acp{AGB} they find is about 5 times higher than those found by other authors \citep[such as][who found just 7 AGBs and did not compute the $R_2$ value for this cluster]{Lagioia2021}, therefore affecting the ratio value.
On the other hand, we find that our $R_2$ ratios also very well agree with other theoretical predictions, which find that $R_2$ should not exceed 0.2 \citep{Cassisi2014}.

Figure~\ref{fig:Ratios} also shows the dependence of \ac{AGB} star occurrences with the initial He-enrichment.
The maximum age at which \ac{AGB} stars appear decreases as the initial He increases. 
However, assuming a different set of input physics for our stellar models may change this threshold age.
For instance, including atomic diffusion may change the stars' lifetimes by up to $10-20\%$ \citep[e.g.][]{VandenBerg2002, Dotter2017, Borisov2024}. 
Therefore, the threshold age between the occurrence of \ac{AGB}-manqué and \ac{AGB} will change too.
This prevents us from using our results to discern the presence of \ac{2P} stars in the \ac{AGB} phase with Y$_{ini} < 0.3$.
On the other hand, we can state that in clusters with similar metallicity to those investigated here, the presence of \ac{AGB} stars at ages above 12 Gyr with Y$_{ini} \geq 0.4$ is quite unlikely, as already predicted and discussed by several authors \citep[e.g.,][]{Charbonnel2013, Milone2018, Martins2021}.
Finally, Tab.~\ref{tab:ratios} lists all the $R_1$ and $R_2$ ratio values and the corresponding errors for each metallicity, age, and initial He-enrichment.

%--------------------------------------------------- table
\begin{table*}
    \caption{Globular cluster number ratios ($R_1$ and $R_2$) for different ages and initial He-enrichment. 
    % The number of \ac{AGB} (N$_{AGB}$) is taken from \citet{Lagioia2021}. 
    For the sake of clarity, empty rows of isochrones with the highest Y values are not shown.}  
    \begin{center}
        \resizebox{0.63\textwidth}{!}{
        \begin{tabular}{cccccccccccc}
            \hline
            \\
            \multicolumn{12}{c}{NGC 4590 $-$ \feh = $-2.2$ $-$ Z = 0.0002} \\
            \hline
            Age [Gyr] & 9.0 & 9.5 & 10.0 & 10.5 & 11.0 & 11.5 & 12.0 & 12.5 & 13.0 & 13.5 & 14.0\\
            \hline
            \multicolumn{12}{c}{Y = 0.248}  \\
            $R_1$ & - & 0.39 & 0.39 & 0.38 & 0.37 & 0.35 & 0.33 & 0.31 & 0.31 & 0.31 & 0.43 \\
            $R_2$ & - & 0.18 & 0.17 & 0.15 & 0.14 & 0.14 & 0.14 & 0.13 & 0.13 & 0.13 & 0.05 \\
            \hline
            \multicolumn{12}{c}{Y = 0.26}  \\
            $R_1$ & - & - & - & 0.39 & 0.38 & 0.36 & 0.34 & 0.33 & 0.32 & 0.35 & 0.36 \\
            $R_2$ & - & - & - & 0.13 & 0.13 & 0.13 & 0.13 & 0.13 & 0.13 & 0.11 & 0.09 \\
            \hline
            \multicolumn{12}{c}{Y = 0.27}  \\
            $R_1$ & - & - & 0.41 & 0.40 & 0.39 & 0.39 & 0.38 & 0.37 & 0.38 & 0.36 & 0.38 \\
            $R_2$ & - & - & 0.14 & 0.13 & 0.13 & 0.13 & 0.13 & 0.12 & 0.10 & 0.09 & 0.08 \\
            \hline
            \multicolumn{12}{c}{Y = 0.3}  \\
            $R_1$ & 0.47 & 0.46 & 0.47 & 0.46 & 0.44 & 0.42 & 0.40 & 0.36 & 0.37 & 0.33 & -\\
            $R_2$ & 0.11 & 0.11 & 0.11 & 0.11 & 0.11 & 0.10 & 0.10 & 0.09 & 0.05 & 0.05 & -\\
            \hline
            \multicolumn{12}{c}{Y = 0.33}  \\
            $R_1$ & 0.50 & 0.50 & 0.48 & 0.50 & 0.43 & - & - & - & - & - & -\\
            $R_2$ & 0.11 & 0.11 & 0.11 & 0.11 & 0.09 & - & - & - & - & - & -\\
            \hline
            \multicolumn{12}{c}{Y = 0.37}  \\
            $R_1$ & 0.47 & 0.44 & 0.42 & - & - & - & - & - & - & - & -\\
            $R_2$ & 0.07 & 0.06 & 0.05 & - & - & - & - & - & - & - & -\\
            \hline
            \hline
            \\
            %%%%%%%%% NGC 6752
            \multicolumn{12}{c}{NGC 6752 $-$ \feh = $-1.53$ $-$ Z = 0.0009} \\
            \hline
            Age [Gyr] & 9.0 & 9.5 & 10.0 & 10.5 & 11.0 & 11.5 & 12.0 & 12.5 & 13.0 & 13.5 & 14.0\\
            \hline
            \multicolumn{12}{c}{Y = 0.248415}  \\
            $R_1$ & 0.36 & 0.33 & 0.36 & 0.35 & 0.38 & 0.38 & 0.36 & 0.38 & 0.38 & 0.36 & 0.38 \\
            $R_2$ & 0.16 & 0.17 & 0.17 & 0.17 & 0.16 & 0.16 & 0.17 & 0.15 & 0.15 & 0.15 & 0.12 \\
            \hline
            \multicolumn{12}{c}{Y = 0.27}  \\
            $R_1$ & 0.45 & 0.45 & 0.43 & 0.40 & 0.37 & 0.40 & 0.40 & 0.38 & 0.35 & 0.31 & -\\
            $R_2$ & 0.11 & 0.12 & 0.14 & 0.14 & 0.14 & 0.13 & 0.12 & 0.11 & 0.09 & 0.05 & -\\
            \hline
            \multicolumn{12}{c}{Y = 0.3}  \\
            $R_1$ & 0.47 & 0.48 & 0.49 & 0.48 & 0.51 & 0.46 & 0.40 & 0.35 & 0.28 & - & -\\
            $R_2$ & 0.10 & 0.10 & 0.10 & 0.10 & 0.13 & 0.11 & 0.07 & 0.04 & 0.02 & - & -\\
            \hline
            \hline
            \\
            %%%%%%%%% NGC 2808
            \multicolumn{12}{c}{NGC 2808 $-$ \feh = $-1.15$ $-$ Z = 0.002} \\
            \hline
            Age [Gyr] & 9.0 & 9.5 & 10.0 & 10.5 & 11.0 & 11.5 & 12.0 & 12.5 & 13.0 & 13.5 & 14.0\\
            \hline
            \multicolumn{12}{c}{Y = 0.249}  \\
            $R_1$ & 0.45 & 0.45 & 0.45 & 0.43 & 0.42 & 0.43 & 0.42 & 0.42 & 0.42 & - & -\\
            $R_2$ & 0.12 & 0.11 & 0.11 & 0.10 & 0.10 & 0.11 & 0.11 & 0.11 & 0.11 & - & -\\
            \hline
            \multicolumn{12}{c}{Y = 0.26}  \\
            $R_1$ & 0.46 & 0.47 & 0.44 & 0.44 & 0.44 & 0.45 & 0.45 & 0.46 & 0.32 & 0.29 & 0.22 \\
            $R_2$ & 0.11 & 0.10 & 0.10 & 0.10 & 0.10 & 0.11 & 0.11 & 0.15 & 0.06 & 0.05 & 0.04 \\
            \hline
            \multicolumn{12}{c}{Y = 0.27}  \\
            $R_1$ & - & 0.44 & 0.46 & 0.45 & 0.47 & 0.49 & 0.45 & 0.42 & 0.38 & 0.31 & 0.24 \\
            $R_2$ & - & 0.10 & 0.10 & 0.10 & 0.10 & 0.10 & 0.12 & 0.11 & 0.10 & 0.05 & 0.03 \\
            \hline
            \multicolumn{12}{c}{Y = 0.3}  \\
            $R_1$ & 0.49 & 0.50 & 0.50 & 0.50 & 0.49 & 0.49 & - & - & - & - & -\\
            $R_2$ & 0.09 & 0.10 & 0.10 & 0.10 & 0.10 & 0.10 & - & - & - & - & -\\
            \hline
            \multicolumn{12}{c}{Y = 0.33}  \\
            $R_1$ & 0.53 & 0.54 & 0.44 & - & - & - & - & - & - & - & -\\
            $R_2$ & 0.09 & 0.09 & 0.07 & - & - & - & - & - & - & - & -\\
            \hline
            \multicolumn{12}{c}{Y = 0.37}  \\
            $R_1$ & 0.40 & - & - & - & - & - & - & - & - & - & -\\
            $R_2$ & 0.04 & - & - & - & - & - & - & - & - & - & -\\
            \hline
            \hline
            \\
            %%%%%%%%% NGC 6441
            \multicolumn{12}{c}{NGC 6441 $-$ \feh = -0.5 $-$ Z = 0.008} \\
            \hline
            Age [Gyr] & 9.0 & 9.5 & 10.0 & 10.5 & 11.0 & 11.5 & 12.0 & 12.5 & 13.0 & 13.5 & 14.0\\
            \hline
            \multicolumn{12}{c}{Y = 0.255}  \\
            $R_1$ & 0.36 & 0.35 & 0.35 & 0.35 & 0.36 & 0.37 & 0.38 & 0.39 & 0.41 & 0.43 & 0.44\\
            $R_2$ & 0.10 & 0.12 & 0.13 & 0.13 & 0.13 & 0.12 & 0.12 & 0.12 & 0.12 & 0.12 & 0.12\\
            \hline
            \multicolumn{12}{c}{Y = 0.27}  \\
            $R_1$ & 0.41 & 0.41 & 0.42 & 0.42 & 0.42 & 0.45 & 0.47 & 0.47 & 0.47 & 0.45 & -\\
            $R_2$ & 0.13 & 0.13 & 0.14 & 0.15 & 0.16 & 0.15 & 0.15 & 0.14 & 0.12 & 0.11 & -\\
            \hline
            \multicolumn{12}{c}{Y = 0.3}  \\
            $R_1$ & - & 0.49 & 0.50 & 0.50 & 0.50 & 0.51 & 0.50 & 0.51 & 0.42 & 0.39 & -\\
            $R_2$ & - & 0.10 & 0.10 & 0.10 & 0.10 & 0.10 & 0.10 & 0.10 & 0.08 & 0.07 & -\\
            \hline
            \multicolumn{12}{c}{Y = 0.33}  \\
            $R_1$ & 0.51 & 0.53 & 0.54 & 0.56 & 0.57 & 0.58 & 0.51 & - & - & - & -\\
            $R_2$ & 0.08 & 0.09 & 0.09 & 0.09 & 0.09 & 0.10 & 0.09 & - & - & - & -\\
            \hline
            \hline
        \end{tabular}
         }
    \end{center}
    \label{tab:ratios} 
\end{table*}

\section{Conclusions}
\label{sec:concl}

In this paper, we adopted and extended stellar evolution tracks from \citet{Charbonnel2016}, \citet{Chantereau2017}, and \citet{Martins2021}
such as to include lower stellar masses and to fully process late stages of evolution up to the end of the \ac{TP-AGB} phase.
We described the methodology adopted to build morphologically accurate isochrones in the presence of transitions from tracks with a standard evolutionary path through the HB and the AGB, to tracks which lack either the AGB, or both the HB and the AGB. 
We presented the resulting isochrones for various metallicities, initial He-contents, and ages (spanning from 9 to 14 Gyr).
The stellar tracks and the new isochrones are available in the CDS database.

For each isochrone, we computed the ratios between the number of AGB stars and the number of RGB stars on one hand and of HB stars on the other. 
We find a good agreement with observed counts within 1$\sigma$ for most cases.
That means that with the parameters and initial physical processes chosen, our models predict reasonable lifetimes for the different post-RGB evolutionary phases.
Varying physical processes, such as overshooting, atomic diffusion, mass loss on the RGB, or rotation, might modify the resulting number ratios, but this should not affect our conclusions on the differential importance of He (as shown in Appendix~\ref{sec:app_comp}).

This work is the first methodological paper of a series that aims to study and characterize remote unresolved stellar populations, which will become more readily available thanks to the {\it James Webb}\ and {\it Euclid} space missions.
Complete isochrones are the first step for creating integrated synthetic spectra of simulated populations computed with different initial abundances and stellar physical conditions. 
We have shown that this grid of tracks produces reasonable number ratios of various types of luminous evolved stars, which is a prerequisite for reasonable integrated light fluxes.
The set of tracks and the isochrones presented and discussed here will be a reference set to which we will compare results for stellar evolution models with modified assumptions, such as updated initial chemical compositions, modified surface boundary conditions, or the presence of various internal mixing processes.

\begin{acknowledgements}
The authors acknowledge Christian Boily, Pierre-Alain Duc, Fabrice Martins, Olivier Richard, and Ferreol Soulez for the fruitful discussions that led to the development of the project. 
This work was supported by the Agence Nationale de la Recherche grant POPSYCLE number ANR-19-CE31-0022. 
TD is supported by the European Union, ChETEC-INFRA, project no. 101008324.
CC thanks the Swiss National Science Foundation (SNF; Project 200021$-$212160).
VB, PC, LM are financed in part by the Coordenação de Aperfeiçoamento de Pessoal de Nível Superior – Brazil (CAPES) – Finance Code 88887.580690/2020-00, the Conselho Nacional de Desenvolvimento Cient\'ifico e Tecnol\'ogico (CNPq) under the grants 200928/2022-8, 310555/2021-3, and 307115/2021-6, and by 
Funda\c{c}\~{a}o de Amparo \`{a} Pesquisa do Estado de S\~{a}o Paulo (FAPESP) process numbers 2021/08813-7 and 2022/03703-1. CG has received funding from the European Research Council (ERC) under the European Union’s Horizon 2020 research and innovation programme (grant agreement No 833925, project STAREX).

\end{acknowledgements}

\bibliographystyle{aa} % style aa.bst
\bibliography{biblio} % your references Yourfile.bib
\appendix

\section{Grid comparisons}
\label{sec:app_comp}

\begin{figure}
    \centering
    \includegraphics[width=\columnwidth]{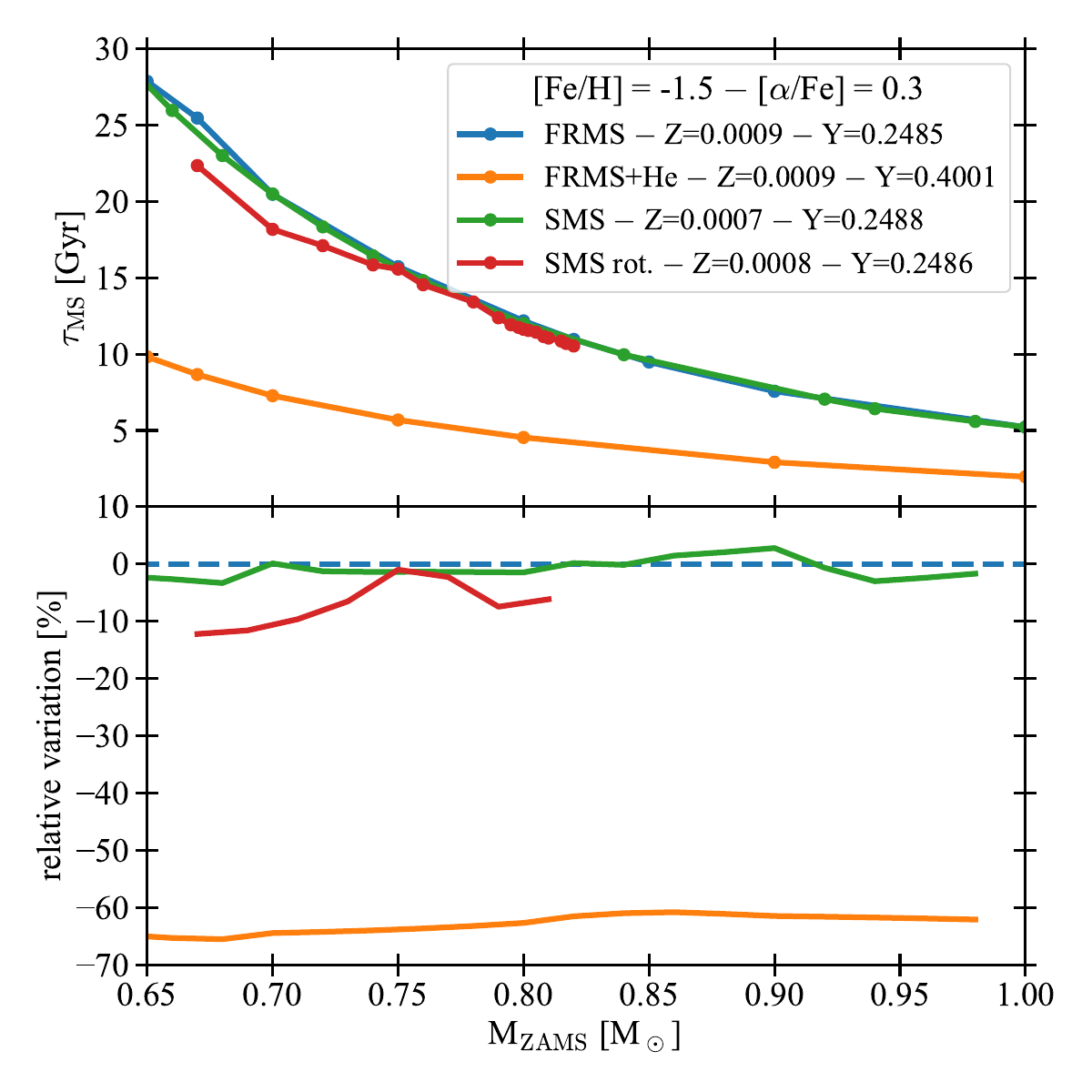}
    \caption{The top panel shows the comparison of the MS lifetimes ($\tau_\mathrm{MS}$) versus the initial mass (M$_\mathrm{ZAMS}$) for tracks computed with different initial abundances, but the same \feh. The bottom panel shows the relative variation of the MS lifetime with respect to the FRMS scenario (i.e., $100 \times \frac{\tau_\mathrm{MS} -\tau_\mathrm{MS}(FRMS)}{\tau_\mathrm{MS}(FRMS)}$), which is indicated by the horizontal dashed line. The points indicate the available tracks.}
    \label{fig:lifetime_comp}
\end{figure}

Here, we compare the properties of our tracks with similar grids computed by other authors. This comparison shows the different lifetimes and number ratios between sets of stellar tracks computed 
with different physical processes and initial chemical abundance.

\subsection{On the impact of different formation scenarios}

Fig.~\ref{fig:lifetime_comp} shows the \ac{MS}-lifetimes vs. the initial mass (M$_\mathrm{ZAMS}$), comparing models from different grids with \feh\ = $-1.5$, all representative of the globular cluster NGC 6752 populations.
For this comparison, we adopted our model computed with standard Y (labeled FRMS) and with enhanced helium (Y = 0.4, grid FRMS+He). 
We also adopted new models (SMS) computed in the super-massive stars framework, which will be published in a forthcoming paper.
These tracks are computed with standard helium, Y = 0.248, without atomic diffusion and rotation, but with thermohaline mixing on the \ac{RGB} phase.
Stellar tracks are computed from 0.2 to 1~\Msun\ and are computed until they reach an age of 15~Gyr.
The last grid of models (labeled SMS rot. in the figure) is taken from new models computed in \citet{Borisov2024}.
These models are also computed in the SMS scenario, with a \feh\ = $-1.5$ and \afe\ = $+0.3$. 
The tracks include rotation-induced processes, atomic diffusion, penetrative convection, parametric turbulence, and additional viscosity.

This analysis shows that the effect of detailed abundances of elements that count in the metal mass fraction Z (associated with different self-enrichment scenarios) has a minimal impact with respect to that of He. 
Specifically, we observe a small relative variation of a few percent when comparing the MS lifetimes of tracks from the FRMS and SMS grids.
On the other hand, the variation changes significantly (more than 60\%) when comparing them with helium-enriched tracks.
From the comparison with the set SMS rot., we find that rotation and atomic diffusion may contribute to variations up to 12\% (in this case).
Again, helium enrichment has a stronger impact on evolution.

\begin{figure}
    \centering
    \includegraphics[width=\columnwidth]{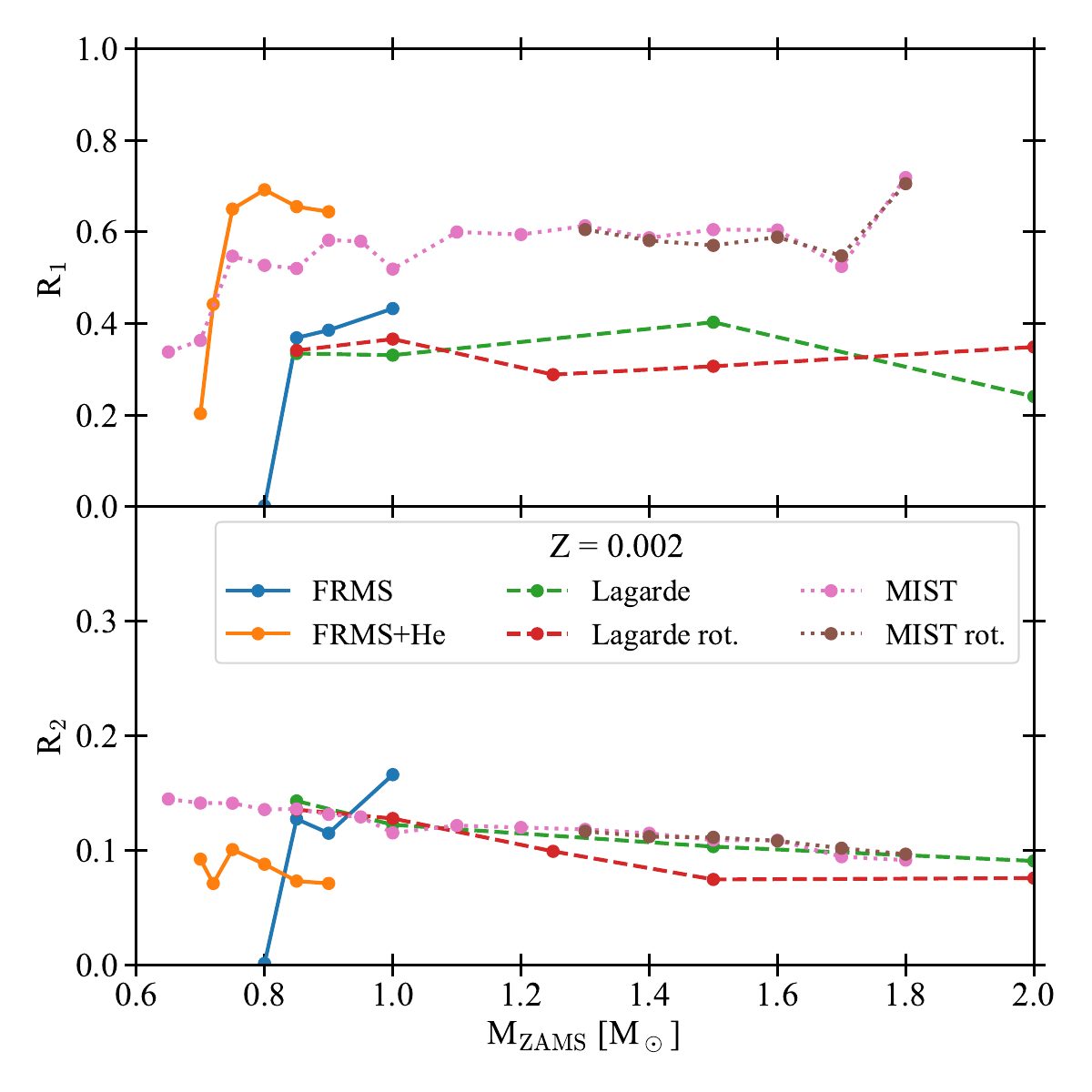}
    \caption{The top and bottom panels compare the R$_1$ and R$_2$ ratios for different grids of tracks, respectively. All the values are computed from tracks with Z = 0.002. Continuous lines indicate tracks presented in this work. The FMRS tracks have a standard He (Y = 0.249) and \afe\ = +0.3. Tracks FMRS+He are computed with Y = 0.4. Dashed lines show models from \citet{Lagarde2012} with (Lagarde rot.) and without rotation (Lagarde). All these tracks have Y = 0.25 and \afe\ = 0. Dotted lines show the \textsc{mist} models from \citet{Choi2016}, with (MIST rot.) and without rotation (MIST). The models are computed with Y = 0.252 and \afe\ = 0.}
    \label{fig:ratios_comp}
\end{figure}

\subsection{On the differential impact of He on number ratios}
We compared the number ratios obtained from our models with those obtained from the grids created by \citet{Lagarde2012} and \citet{Choi2016}  to understand how different processes impact them.
We chose these two grids because they are similar to our sets of tracks, as they compute the full evolution from the ZAMS to the TP-AGB phases. In this analysis, we focused on comparing the impact of rotation with that of helium.

Fig.~\ref{fig:ratios_comp} shows the comparison between the number ratios R$_1$ and R$_2$ with respect to the initial mass of stars. 
The number ratios are derived from the evolutionary lifetimes of each track using the same critical points selection method, described in Sec.~\ref{subsec:cp}.
The comparison reveals that the impact of stellar rotation on number ratios is relatively small. 
In fact, by analyzing the largest variations for the same initial mass, it was found that non-rotating and rotating tracks differ by less than 0.1 and 0.04 for R$_1$ and R$_2$, respectively, for the \citet{Lagarde2012} grids. Similarly, for the \citet{Choi2016} tracks, the difference is less than 0.04 and 0.02 for R$_1$ and R$_2$, respectively.
On the other hand, a He-enrichment of 0.15 in the initial composition (FRMS models) leads to variations greater than 0.2 and about 0.06 in the R$_1$ and R$_2$ values, respectively.
It is worth noticing that the largest rotation-related variation for the MIST and Lagarde grids is seen for models with M$_\mathrm{ZAMS} > 1.4~\Msun$. 
These models reach the AGB phase at around 2~Gyr, which is much younger than the age of globular clusters. 
Finally, according to the models presented in \citet{Lagarde2012}, the difference between non-rotating and rotating tracks is negligible for stars with a mass of 0.85 solar masses. 
Unfortunately, \citet{Lagarde2012} stellar track database\footnote{\url{https://cdsarc.cds.unistra.fr/viz-bin/cat?J/A+A/543/A108}} does not provide models with masses below 0.85~\Msun, focusing instead on the intermediate-mass range.
On the other hand, the MIST\footnote{\url{https://waps.cfa.harvard.edu/MIST/index.html}} database does not contain rotation models for stars with masses below 1.3 solar masses \citep{Choi2016}.

\section{Isochrones building methods}
\label{sec:app_criteria}

\begin{figure*}
    \centering
    \includegraphics[width=\textwidth]{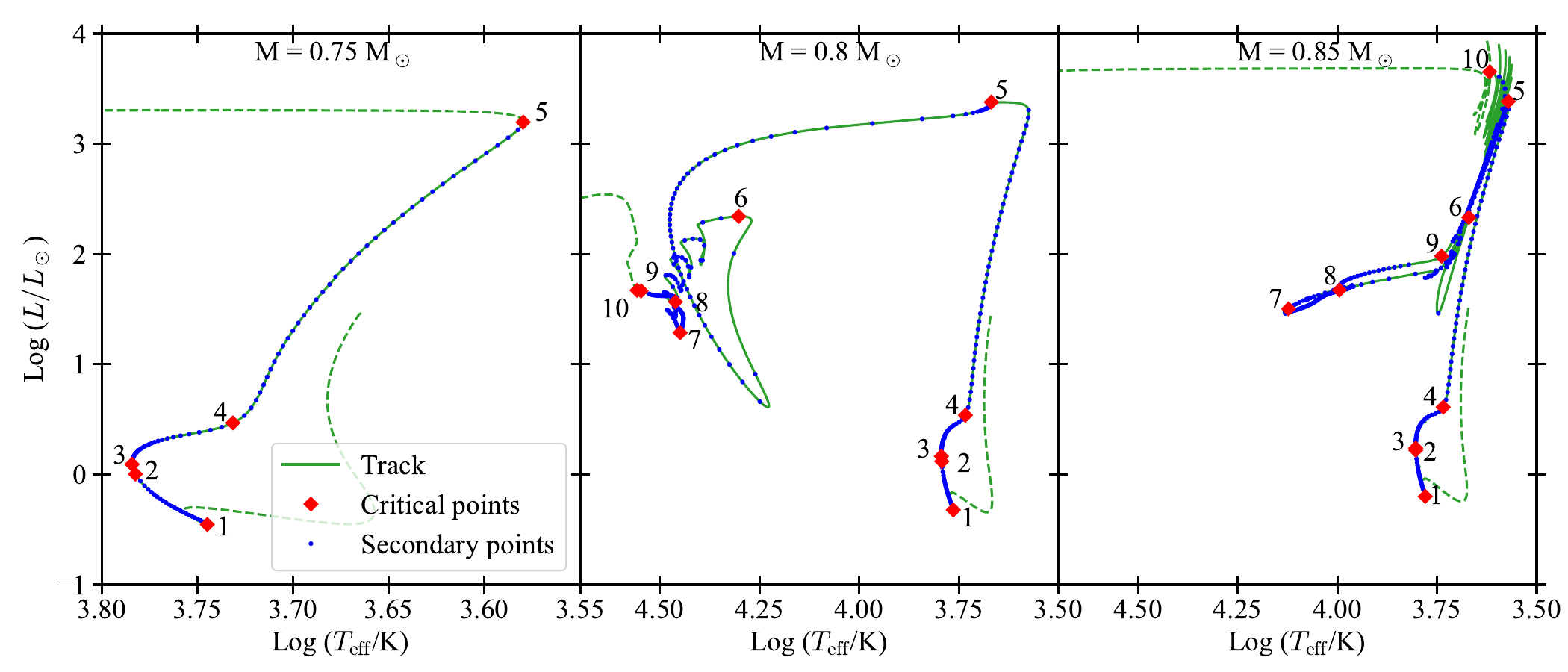}
    \caption{\ac{HR} diagram of selected tracks with 0.75, 0.8, and 0.85~\Msun\ in the left-hand, middle, and right-hand panels, respectively. The tracks are computed with Z = 0.002 and Y = 0.249. The continuous lines show the evolutionary phases we include in the isochrones computation, while dashed lines indicate the pre-\ac{MS} and the planetary nebula phases (excluded from the computations). Red diamonds and blue circles show the critical points (marked by a number) and the secondary interpolated points, respectively.}
    \label{fig:HR_cp}
\end{figure*}

\begin{figure*}
    \centering
    \includegraphics[width=\textwidth]{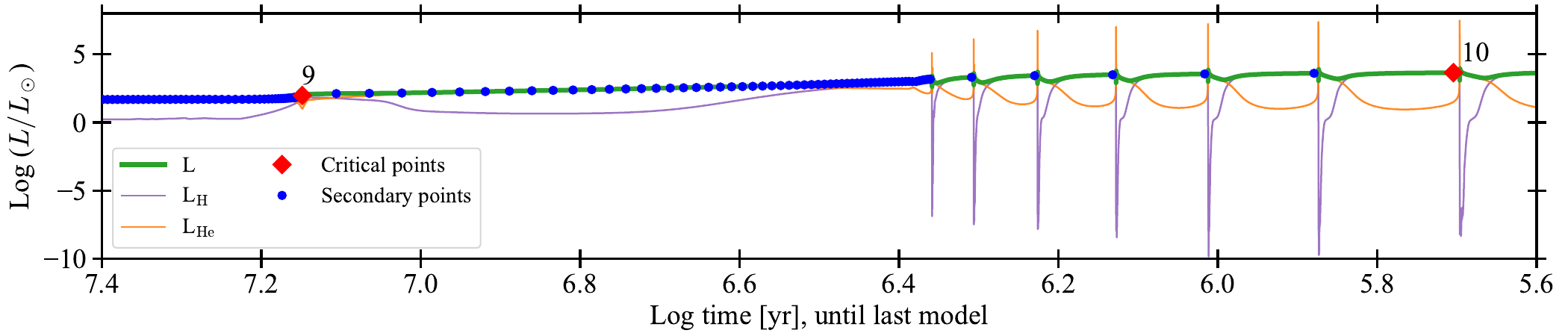}
    \caption{Evolution of the luminosity vs. time during the \ac{AGB} phase of a $M_\mathrm{ZAMS} = 0.85~\Msun$ track with Z = 0.002 and Y = 0.249. Red and blue points indicate the critical and secondary points, respectively. The thin purple and orange lines show the evolution of the luminosity provided by H- and He-burning, respectively.}
    \label{fig:Kipp_AGB_cp}
\end{figure*} 

The method we adopted to build our new isochrones consists of three steps.
The first is the selection of critical points and the construction of reduced stellar track tables.
The second consists of the creation of the `fake' track when needed.
The last one regards the isochrone building performed with \textsc{syclist}. 
We present all the important details in the following sections.

\subsection{1st-step -- New critical points selection}
\label{subsec:cp}

\begin{figure*}
    \centering
    \includegraphics[width=\textwidth]{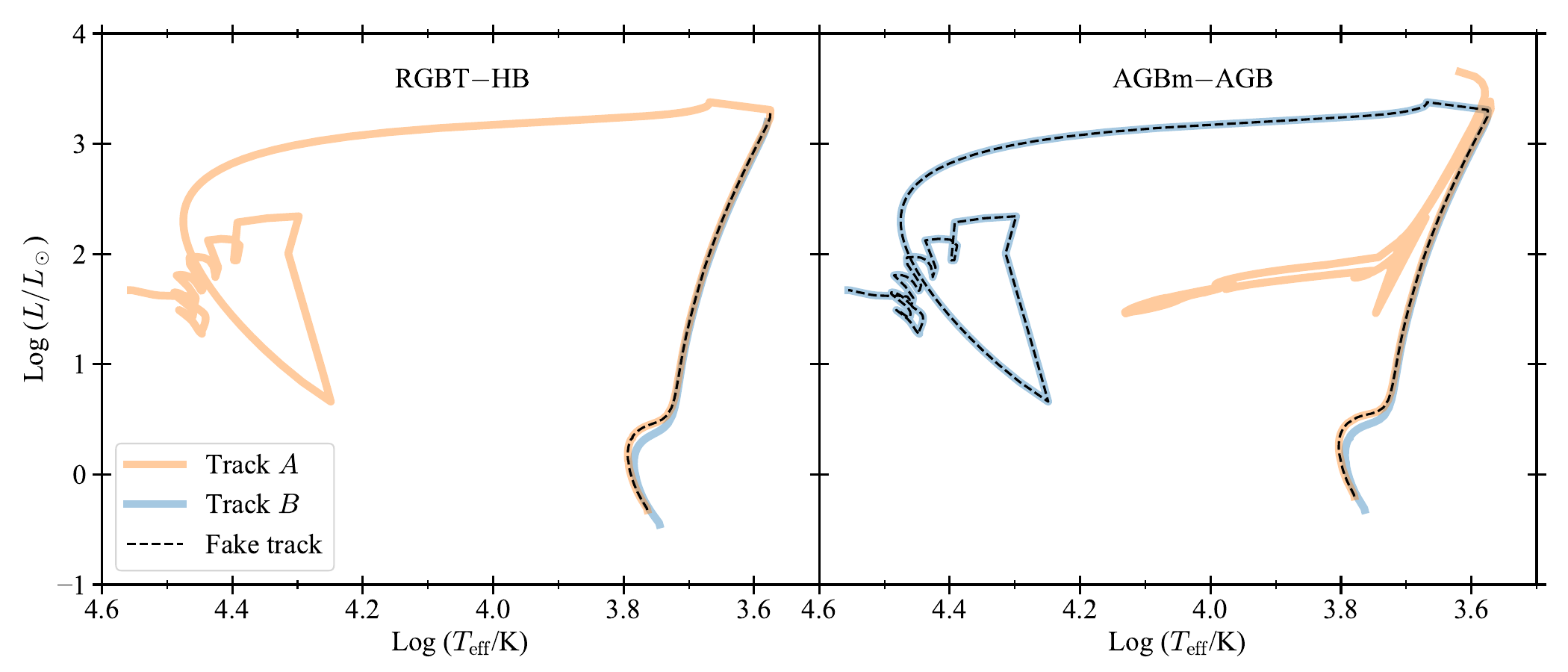}
    \caption{Two examples of `fake' tracks (dashed black lines) created to avoid the interpolation between tracks with different evolutionary paths. The orange line shows the track with the mass above the transition (track $A$), and the blue line indicates the track with the mass below the transition (track $B$). The left-hand panel shows the fake track created for the transition RGBT-HB (enclosed by tracks with 0.75 and 0.8~\Msun), while the right-hand panel shows the `fake' track for the AGBm-AGB transition (enclosed by tracks with 0.8 and 0.85~\Msun). All the tracks are computed with Z = 0.002 and Y = 0.249. See details on the `fake' track creations in the text. 
    }
    \label{fig:HR_faketrack}
\end{figure*}

From each evolutionary track, we create reduced tables containing 500 points selected to describe the whole evolution from the \ac{ZAMS} to the luminosity tip of the \ac{RGB} or \ac{AGB} phases. 
Firstly, we determine the so-called critical points (cp) that split the different evolutionary phases \citep[see also][]{Ekstrom2012, Chantereau2015}. 
Our selection of the critical points is optimized to sample all the characteristic morphological features of low-mass star tracks in the \ac{HR} diagram, and represent an updated version of that presented by \citet{Chantereau2015}.
For stellar tracks that end their evolution after the \ac{RGB} tip, we select five cps, while for those that ignite helium and evolve in the \ac{HB}, we select ten cps.
The critical points are selected as follows:
\begin{enumerate}
    \item beginning of the main-sequence (i.e., \ac{ZAMS}), selected when in the star's centre $X(\mathrm{H_{ini}}) - X(\mathrm{H}) < 0.003$, where $X(\mathrm{H_{ini}})$ is the initial hydrogen abundance in mass fraction;
    \item \ac{MS} turn-off point, when central $X(\mathrm{H}) = 0.05$;
    \item \ac{MS}-end, when central $X(\mathrm{H}) < 10^{-3}$;
    \item bottom of the \ac{RGB} phase, which is defined as the moment when the mass of the convective envelope is more than 25\% of the star's total mass;
    \item tip of the \ac{RGB}, i.e. most luminous point of the \ac{RGB} phase;
    \item first helium luminosity peak after He-flash;
    \item beginning of the \ac{CHeB} phase, when central $X(\mathrm{He}) < 0.98$;
    \item end of the \ac{CHeB} phase, when central $X(\mathrm{He}) < 0.05$;
    \item bottom (beginning) of the early-\ac{AGB} phase, when two burning shells co-exist.
    \item Maximum luminosity (tip) of the \ac{TP-AGB} phase.
\end{enumerate}
Figure~\ref{fig:HR_cp} shows the critical points selected for three tracks with different evolutionary paths, as discussed in Section~\ref{subsec:paths}.
The selection of the critical points ensures that the secondary points (selected as described below) are evolutionary equivalent between all tracks.

To determine the secondary equivalent points, we interpolate all intervals between two critical points as follows:
\begin{itemize}
    \item We use 25 points to sample both [cp1-cp2] and [cp2-cp3] intervals, with linear steps in central $X(\mathrm{H})$ abundance.
    \item 40 points to interpolate the interval [cp3-cp4], with linear steps in time.
    \item 40 points to interpolate between [cp4-cp5], in equal steps of log $L$.
    \item We select 80 and 65 points in the intervals [cp5-cp6] and [cp6-cp7], respectively. We use equal steps in time to perform the interpolation.
    \item For the interval [cp7-cp8] (i.e., \ac{CHeB} phase), we use 50 points with equal steps in the cumulative sum of the central He abundance, $X(\mathrm{He})$. This assures a monotonic trend of the interpolating variable\footnote{The not monotonic decrease of $X(\mathrm{He})$ could be due to thermal breath pulses or potential numerical spikes.}.
    \item From the end of the \ac{CHeB} to the beginning of the \ac{AGB} phase (i.e., interval [cp8-cp9]), we select 75 points with linear steps in time.
    \item Finally, for the interval [cp9-cp10] that corresponds to the \ac{AGB} phase, we use 100 points with equal steps of the cumulative sum of log $L$. 
    To simplify the difficult interpolation of the \ac{TP-AGB} phase, we pick only the most luminous points of each inter-pulse quiescent phase in our collection of secondary points, as shown in Figure~\ref{fig:Kipp_AGB_cp}. 
    Such a simplified approach has been used by other authors in the literature \citep[such as][]{Bertelli1990, Mouhcine2002, Marigo2008}, and is opposed to other approaches which instead add the thermal pulses in isochrones a posteriori \citep{Marigo2017}.
\end{itemize}
For the stars that do not go through the He-burning phase, the last line of their evolution, corresponding to the fifth cp (i.e., \ac{RGB}-tip), is repeated up to line 500, so we keep the same size for all the reduced tables.

\subsection{2nd-step -- `Fake' star creation process}
\label{subsec:fakestar}

The second step for computing smooth isochrones consists of the creation of `fake' stellar evolution tracks. 
These tracks are created to cure the quick change of behavior between tracks that go to the planetary nebula phase just after the \ac{RGB} tip and those that experience the He-flash (we refer to this transition as RGBT-HB), and the transition between tracks that evolve as \ac{AGB}-manqué and those that go to the standard \ac{AGB} phase (transition name AGBm-AGB).
In both cases, the mass interval in which the change of behavior happens is $\Delta M < 0.05~\Msun$.
An example is shown in Figure~\ref{fig:HR_cp}.

The main idea behind the `fake' track method is to replace the real transition (which could be known only by computing a denser grid of tracks) with an abrupt transition that occurs at an initial mass almost equal to the initial mass of one of the available tracks.
The `fake' track shares the evolutionary properties with each of the two nearest tracks, which enclose the transition. 
For the sake of simplicity, we use $A$ to refer to the track with the mass above the transition and $B$ to refer to the track with the mass below.
For the stellar properties, we use the subscript to indicate which track we refer to (e.g., $M_A$ is the initial mass of the track above).
For both transitions, we assume that the `fake' track has a mass $M_\mathrm{fake} = M_A - \varepsilon$, where $\varepsilon = 10^{-11}~\Msun$\footnote{It is important to use such a small value for $\varepsilon$ to avoid incorrect interpolation between the tracks. As explained in Section~\ref{subsec:SYCL}, we need to use very tiny $dm$ in the isochrone building to follow the fast evolution after the He-flash, and obtain accurate isochrones.}.

For the first transition (i.e., RGBT-HB), we assume that the `fake' star follows the evolution of the track $A$ until the \ac{RGB} tip.
Then, we cut all the evolution after the tip, thus assuming that the `fake' star has the same post-RGB tip evolution as the track $B$.
All the evolutionary properties (also time) of the `fake' track are exactly the same as the track $A$, except for the initial mass that is smaller by $\varepsilon$.
Things are a bit more complex for the second transition, AGBm-AGB.
As before, the `fake' track follows the evolution of the track $A$ until the \ac{RGB} tip. 
After the tip, we assume that the `fake' star follows the evolution of the track $B$. 
The time is the only `fake' star property that remains the same as that of track $A$ for the whole evolution.
An example of `fake' track reconstruction for the two transitions (RGBT-HB and AGBm-AGB) is shown in Figure~\ref{fig:HR_faketrack}. The Figure also shows the two tracks closer to the transition.

This approach implies that we are assuming that all the tracks between $M_B$ and $M_A$ evolve in the same way as track $B$, and the transition between the two evolutionary paths happens in a $dm = \varepsilon~\Msun$.
Creating and adding the `fake' track to the collection of tracks adopted to build isochrones assures that we do not interpolate between tracks with different paths and, therefore, permits the computation of isochrones with more reliable morphologies, which do not show strange artifacts, like those shown in Figure~\ref{fig:HR_without_fakestar}.

\paragraph{Criterion of AGB selection.}
The criterion chosen in order to detect the occurrence of AGB-manqu\'e evolution is fundamental to properly applying the `fake' track methodology. 
We adopted the following check to detect the AGBm behavior:
1- Check the occurrence of at least two thermal pulses. If they are present, proceed to step 2; otherwise, AGBm is detected.
2- Check if the pulse positions in the \ac{HR} diagram satisfy the following $\Teff < 3.85 $ and $\log L > 3.1$. If yes, standard AGB stars are detected; otherwise, AGBm is detected.
Other criteria based on other stellar properties, such as the envelope mass vs total mass ratio \citep{Charbonnel2013}, or the central degeneracy at the \ac{RGB} tip \citep{Cassisi2014}, may help. Still, they do vary with the initial composition and mass. Therefore, it is difficult to establish a general criterion.

\subsection{3rd-step -- isochrone building}

\textsc{syclist} uses an adaptive mass step ($dm$) for the isochrone building process.
The mass step is iteratively reduced when the difference between two sequential points of the isochrone satisfies one of the following conditions: $\delta \log T_\mathrm{eff} > 0.02$ or $\delta \log L > 0.02$.
When the condition is hit, the code recomputes the new mass point with a reduced $dm$.
A minimum mass step ($dm_{min}$) is imposed to prevent the code from diverging to infinitesimal $dm$, avoiding never-ending computations.
In fast evolutionary phases, such as after the He-flash, a very small mass step is required to follow that phase properly. 
We found that $dm_{min} = 10^{-10}$ is good enough to interpolate the fast phases and to build good isochrones.
We stress here that the choice of an $\varepsilon < dm_{min}$ is critical to minimize the chances of interpolation between a track above the transition and the `fake' one, as described in Section~\ref{subsec:fakestar}.

\begin{figure}
    \centering
    \includegraphics[width=\columnwidth]{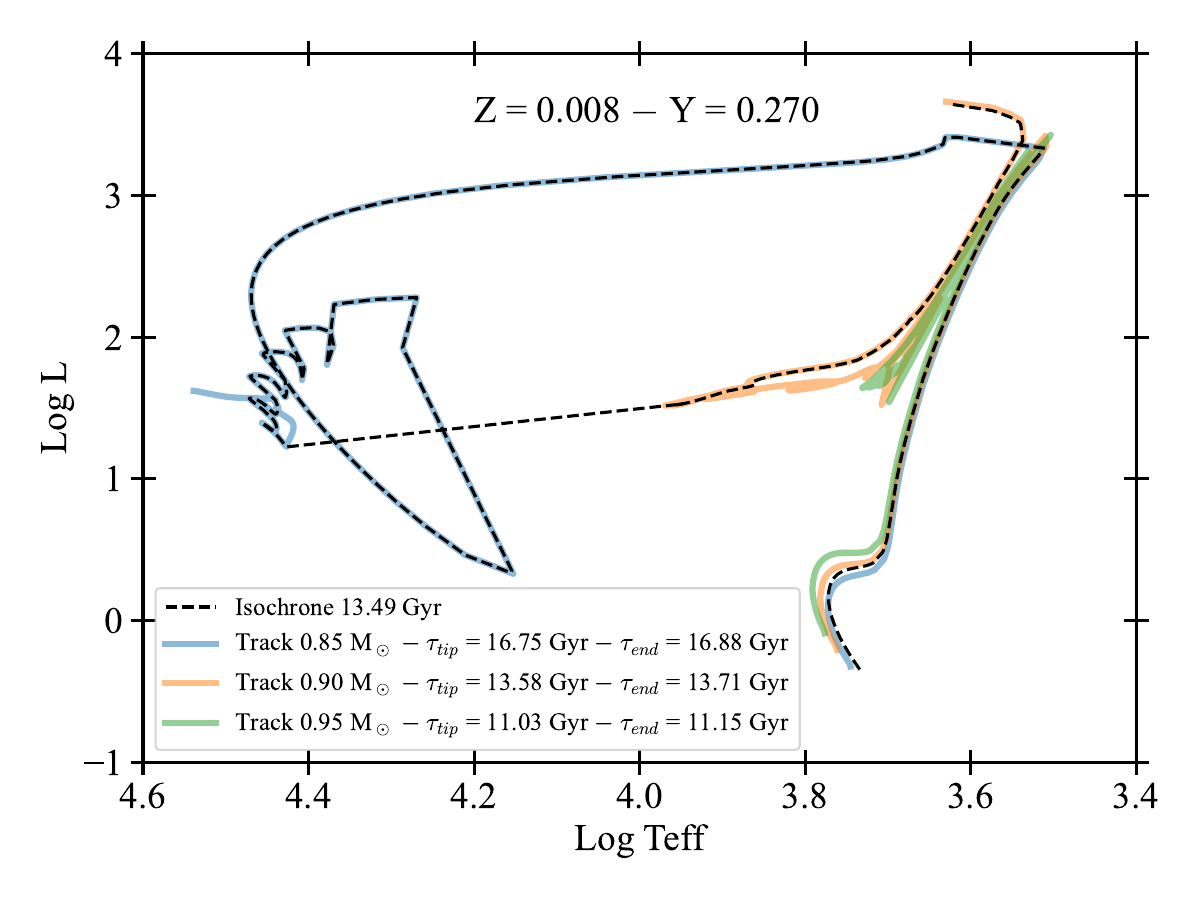}
    \caption{This figure shows an example of the wrong use of the fake track in the interpolation to build an isochrone of 13.5 Gyr (indicated by the dashed black line). The two interpolated tracks are shown in blue and orange, while the correct interpolation should be done between the orange and green tracks. The metallicity is 0.008, and the He content is 0.27.}
    \label{fig:fake_wrong_use}
\end{figure}

\paragraph{Isochrones near the transition masses.}
The `fake' tracks should be used carefully to avoid wrong interpolation in the isochrone building process. 
To properly use `fake' tracks, it is important to check the age at the tip of the \ac{RGB}. 
The `fake' track should be included in the collection of tracks used to compute isochrones only if the isochrone age is enclosed by the ages at the RGB tip of the two tracks with different behaviors.
Figure~\ref{fig:fake_wrong_use} shows a result of the blind usage of `fake' tracks.
In this case, the isochrone in the post-\ac{RGB} part should be interpolated between the two tracks with 0.90 and 0.95~\Msun.
Including the `fake' track, mixed up the interpolation algorithm, and create the wrong interpolation.
Therefore, to obtain accurate isochrones, it is important to use `fake' tracks only when needed.

\end{document}